\documentclass[preprint,review,12pt,sort&compress]{elsarticle}

\usepackage{amssymb}
\usepackage{amsthm}
\usepackage{graphicx}
\usepackage{amsmath}
\usepackage{bm}
\usepackage{stmaryrd}
\usepackage{subfigure}
\usepackage[export]{adjustbox}
\usepackage{tikz}
\usepackage[colorlinks=true]{hyperref}
\usepackage{geometry}
\usepackage{setspace}
\usepackage{longtable}
\usepackage{physics}
\usepackage{mathtools}
\usepackage{algorithm}
\usepackage{algpseudocode}
\usepackage{listings}

\usepackage{xcolor}
\newcommand{\firstreviewer}[1]{{\color{red}#1}}
\renewcommand{\firstreviewer}[1]{#1} 
\newcommand{\secondreviewer}[1]{{\color{blue}#1}}
\renewcommand{\secondreviewer}[1]{#1} 
\newcommand{\removedetails}[2]{#2}

\newcommand{\argmin}[1]{{\arg\min_{#1}\,}}

\newcommand{\dev}[1]{\text{dev}\left(#1\right)}
\renewcommand{\trace}[1]{\text{tr}\left(#1\right)}
\newcommand{\assembleOperator}{\bigwedge}

\algrenewcommand\algorithmicensure{\textbf{Output:}}

\graphicspath{{images/}{.}} 

\journal{}

\makeindex

\begin{document}

\begin{frontmatter}
	
	
	
	\title{A quantum annealing-sequential quadratic programming assisted finite element simulation for  non-linear and history-dependent mechanical problems}
	
	
	\author[label1]{Van-Dung Nguyen\fnref{FNRS}}\ead{vandung.nguyen@uliege.be}
        \author[label1]{Ling Wu}\ead{L.Wu@ulg.ac.be}
	\author[label2]{Fran\c{c}oise Remacle}\ead{fremacle@ulg.ac.be}
	\author[label1]{Ludovic Noels\corref{cor}}\ead{L.Noels@ulg.ac.be}
	
	\cortext[cor]{Corresponding author, Phone: +32 4 366 48 26, Fax: +32 4 366 95 05}
	\address[label1]{Computational \& Multiscale Mechanical of Materials (CM3), \\ Department of Aerospace and Mechanical Engineering, \\University of Li\`ege, \\ Quartier Polytech 1, All\'ee de la D\'ecouverte 9, B-4000 Li\`ege, Belgium}
	\address[label2]{MolSys Research Unit,\\University of Li\`ege, \\ Quartier Agora,
		All\'ee du six Ao\^ut 11, B-4000 Li\`ege, Belgium}
	\fntext[FRNS]{Postdoctoral Researcher at the Belgian National Fund for Scientific Research (FNRS)}
	
	\begin{abstract}
		We propose a framework to solve non-linear and history-dependent mechanical problems based on a hybrid classical computer -- quantum annealer approach. Quantum Computers are anticipated to solve particular operations exponentially faster. The available possible operations are however not as versatile as with a classical computer. However, quantum annealers (QAs) are well suited to evaluate the minimum state of a Hamiltonian quadratic potential.
		Therefore, we reformulate the elasto-plastic finite element problem as a double-minimisation process framed at the structural scale using the variational updates formulation.
		In order to comply with the expected quadratic nature of the Hamiltonian, the resulting non-linear minimisation problems are iteratively solved with the suggested Quantum Annealing-assisted Sequential Quadratic Programming (QA-SQP): a sequence of minimising quadratic problems is performed by approximating the objective function by a quadratic Taylor's series. Each quadratic minimisation problem of continuous variables is then transformed into a binary quadratic problem. This binary quadratic minimisation problem can be solved on quantum annealing hardware such as the D-Wave\footnote{D-Wave trademarks and registered trademarks used herein include D-Wave and Ocean.} system. The applicability of the proposed framework is demonstrated with one- and two-dimensional elasto-plastic numerical benchmarks. The current work provides a pathway of performing general non-linear finite element simulations assisted by quantum computing.
\end{abstract}
	
\begin{keyword}
		Quantum computing \sep Quantum annealing \sep Variational formulation \sep Finite element \sep Elasto-plasticity
\end{keyword}
	
\end{frontmatter}


\section{Introduction} \label{sec:intro}

Continuum mechanics represents the behaviour of solids and fluids through a system of complex non-linear Partial Differential Equations (PDEs).
Whilst analytical solutions can be found for some simple cases, finite element (FE) method is a standard tool widely used to find approximate solutions, which is based on the discretisation of the continuum domain into a finite element mesh consisting of a finite number of elements and nodes.
The physical field is thus approximated inside each element using a limited number of unknowns, which are often the nodal values, and which interpolate the fields through their associated shape functions.
Such a procedure allows turning the continuous PDEs into a discrete form with a finite number of continuous degrees of freedom, and can be solved by different numerical techniques \cite{wriggers2008nonlinear}.
Only in a very limited situation, in which the underlying material behaviour is linear elastic, yields linear governing PDEs.
Therefore, high computational resources are needed when the computation domain is very large, because of the very large resulting number of degrees of freedom, and/or when complex physical behaviours such as plasticity, damage and fracture are involved, because of the highly non-linear nature of the resulting PDEs.

Quantum computing is a rapidly developing field of computer science that uses the principles of quantum mechanics. Unlike classical computers, which store and manipulate information using bits represented by binary values (0s and 1s), quantum computers use quantum bits or so--called qubits, which can take values of 0 or 1, but which can also exist in a superposition of both 0 and 1 at the same time. Taking advantage of quantum effects allows quantum computers to potentially solve particular kinds of problems that are intractable for classical computers \cite{arute2019quantum}, see also the discussions in \cite{pednaud2019IBM,castelvecchi2023nature}. 

There exist two main quantum implementations.
On the one hand, gate-based quantum computers rely on so-called quantum gates to manipulate qubits, taking advantage of quantum superposition and entanglement to perform computations \cite{nielsen_chuang_2010}. Quantum gates are analogous to classical logic gates, but operate on qubits instead of bits as in classical computers. Gate-based quantum computers are currently the most advanced type of quantum computer for universal computations. However, they require extremely precise control over the qubits and their environment, and are currently limited in size by the difficulty of maintaining this control. Waiting for efficient Quantum Error Correction (QECs), gate-based quantum computers are practically of the noisy intermediate-scale quantum class computers \cite{NAP25196}.
On the other hand, quantum annealers embracing quantum annealing technologies \cite{kadowaki1998quantum,brooke1999quantum,tanaka2017quantum} are designed to solve optimisation problems by starting with a system of qubits in a simple state and then slowly modifying their interactions or entanglements until they settle into a state that represents the optimal solution of the problem. Quantum annealers are typically used to solve combinatorial optimisation problems and are not as versatile as gate-based quantum computers, which can run a wider range of algorithms.

Application of quantum computing in finite element simulations is a tempting target, see the review by \cite{Balducci2022}. A typical numerical technique, such as the finite element method, follows an iterative procedure in which a linear system needs to be solved at every iteration, and its solution is used to update the current solution until a terminating criterion is met \cite{wriggers2008nonlinear}.
In this context, the exciting recent advance in the development of quantum algorithms in gate-based quantum computers for solving a large system of linear equations allows potentially solving large FE problems on quantum computers.
Indeed, quantum linear solvers \cite{childs2017quantum,harrow2009quantum} could provide an exponential quantum speed-up over classical algorithms when solving linear systems. The system of linear equations arising from the FE method is sparse, thus satisfying an important requirement of the exponential quantum speed-up \cite{montanaro2016quantum}.
However, there are still many challenges that need to be overcome before it can be effectively used for FE simulations.
Firstly, quantum computers are still in their early stages as the current quantum hardware is sensitive to noise and has high error rates, in particular for gate-based quantum computers, which makes it difficult to implement and execute quantum algorithms accurately and efficiently. Secondly, the existing FE algorithms are designed to run on classical computers: the resolution of the system of linear equations is only one simple step in the workflows, and the remaining operations are not readily adaptable to quantum computing architectures.
However, many of the steps in a typical finite element simulation can rely on classical computations, so integrating quantum algorithms into classical computing could be considered \cite{cerezo2021variational}, requiring the development of en efficient hybrid classical computer -- quantum annealer approach that can effectively use both quantum and classical resources.

Quantum annealers can be used to solve the system of linear equations, but this requires reformulating the problem to account for continuous variables.
A general rule to apply quantum annealers is first to define the solution of this linear system as minimisation of a quadratic objective function, then to approximate each continuous variable by a discrete representation using the radix-2 approximation, leading to a binary quadratic form, which is an objective function compatible with quantum annealers \cite{o2016toq,borle2019analyzing,date2021qubo,Criado_2023}.
Each binary variable whose value can be either 0 or 1 is represented by a qubit in the quantum annealers. At the end of the annealing process being programmed to minimise the binarised objective function, the measurement of the qubits provides the optimised solution, offering a speed advantage over classical methods.
However, the solution precision in this so-called direct annealing approach is limited, on the one hand by the number of qubits used because of the radix-2 encoding, and on the other hand by the fact that the ground state solution of the quantum system is not guaranteed due to noise and errors in the quantum annealers.
These issues can be improved using the so-called iterative annealing approach in which a series of minimisation problems are performed by quantum annealers to get better radix-2 approximation \cite{conley2023quantum} with a lower number of qubits per degree of freedom until a terminating criterion is met. Such an iterative procedure generally does not require the correct ground state of the quantum system to be reached, but only to reach a better solution after each iteration. 

Quantum annealers were directly considered in the context of the FE method. The box algorithm  proposed in \cite{Srivastava2019} consists in iteratively mapping each degree of freedom with 8 discrete values using 3 qubits and performing the quantum annealing to find the optimised solution until the desired precision is met.
An iterative FEqa formulation was proposed in \cite{RAISUDDIN2022115014}, in which each unknown is mapped into a singe qubit and it was shown that the FEqa has clear advantages in computational time when considering the quantum annealing instead of considering simulated annealing.
An iterative algorithm to solve the elliptic problems on the D-Wave quantum annealing device was proposed \cite{conley2023quantum}, in which a fixed number of qubits is used for each degree of freedom parameterised by a radix-2 approximation, the range of each each degree of freedom being adjusted after each iteration. Nevertheless, these methods are considered only in the context of the linear elasticity. Quantum annealing was also considered in the context of predicting phase separation with the phase-field method \cite{Endo2022}, in which the potential arising from the physical energies (the interaction --internal-- energy term, the gradient --adjacent-- energy term, and the long-distance energy term) at play has to be minimised. In this approach the unknowns are discretised by 4 binary variables, without continuous representation.

The present work focuses on developing a hybrid classical computer -- quantum annealer framework for general non-linear and history-dependent finite element simulations. First, we formulate the elasto-plastic solid mechanics boundary value problem as non-linear minimisation problems using the variational updates formulation \cite{ortiz1999variational,radovitzky1999error,weinberg2006variational}, leading to a so-called double-minimisation problem.
Not only the unknowns field, \emph{e.g.} displacements, but also the internal states, \emph{e.g.} plastic deformation gradients, are obtained through minimisation processes within each load increment. Contrarily to what is classically done in finite element codes, in which the material model equations are solved locally at the Gauss point level whilst only the nodal degrees of freedom are solved at the structural scale in a nested way, in order to take advantage of the potential ability of quantum computing to solve large scale problem, this double-minimisation problem is expressed in terms of both the unknown field and internal states at the structural scale, so that only assembly operations have to be performed on the classical hardware.
Then, in order to comply with the requested quadratic nature of the minimisation problems that can be solved by quantum annealers, a Sequential Quadratic Programming (SQP) with quantum annealing --or so-called Quantum Annealing-assisted Sequential Quadratic Programming (QA-SQP)-- is suggested to solve a general non-linear minimisation problem using quantum annealers. Although, in general, a minimisation can be carried out by iteratively carrying out many minimisation problems of a single variable, \emph{e.g.} line search, steepest descent, conjugate gradient methods \emph{etc.} \cite{hager2006survey,ruszczynski2011nonlinear}, in this work, a Sequential Quadratic Programming (SQP) iteratively approximates the objective function using a quadratic Taylor's series and solves a sequence of minimisation quadratic problems \cite{bonnans2006numerical}. SQP is typically used to find the minimum of a non-linear function subject to constraints, such as inequality constraints or equality constraints.
In the developed QA-SQP, the quadratic minimisation problems are solved by quantum annealing using an adaptive binary approximation to transform the continuous quadratic minimisation problems into quadratic binary problems.

The paper is organised as follows. Section \ref{sec:qaSummary} briefly introduces the quantum annealing working principles.
Section \ref{sec:mechaBVP} provides the mathematical foundation of the variational updates formulation in which we detail the minimisation principle of the mechanical boundary value problem in the context of a spatial finite element discretisation.
Section \ref{sec:QA-SQP} introduces the Quantum Annealing-assisted Sequential Quadratic Programming (QA-SQP) and its particularisation to the resolution of the double-minimisation problem formulated in Section \ref{sec:mechaBVP}.
Section \ref{sec:examples} presents numerical benchmarks with the proposed framework. Finally, conclusions are drawn in Section \ref{sec:conclusions}.

\firstreviewer{
\section{Quantum annealing}\label{sec:qaSummary}
Quantum annealing is based on the principle of adiabatic quantum computation, which involves gradually transforming a quantum system from an initial Hamiltonian, in which its ground state, \emph{i.e.} the stationary state of the lowest energy, can be easily prepared, to a final Hamiltonian, in which the problem is encoded and whose ground state provides the solution.

\subsection{Quantum state representation}
The quantum states refer to the possible configurations in which the quantum system can exist.
The key concepts related to the quantum states which make quantum computing more powerful than the classical one consist of the principle of superposition, in which a quantum system can exist in multiple states simultaneously, and of entanglement, in which the states of two or more particles become correlated in such a way that the state of one particle influences the one(s) of the other(s) regardless of the distance between them.
Understanding and manipulating the quantum states are crucial in quantum computing. More comprehensive details about quantum computing can be found in \cite{nielsen_chuang_2010,Abhijith2022}. In this section, we briefly summarise the mathematical representation of quantum states. 

In quantum computing, a quantum bit (so-called qubit) is the fundamental information unit that can be seen as the generalisation of a binary bit in classical computers. A qubit is a two-state quantum-mechanical system whose state can be represented by any normalised linear combination of the two basis states, which are usually denoted as $\ket{0}=[1\,\,0]^T$ \text{ and } $\ket{1}=[0\,\,1]^T$.
The Dirac notation $\ket{\bullet}$ is used to represent a state of a quantum system, as
\begin{eqnarray}\label{eq:qbitState}
	\ket{\phi} =\phi_0 \ket{0} + \phi_1 \ket{1} \,,
\end{eqnarray}
where $\phi_0$ and $\phi_1$ are complex numbers satisfying $\abs{\phi_0}^2 + \abs{\phi_1}^2 = 1$ to ensure that the total probability of being in any of the basis states is equal to 1.\removedetails{ Eq. (\ref{eq:qbitState}) implies that a qubit can be both 0 and 1 at the same time, which is known as superposition, whereas a classical bit can only be 0 or 1, never both.}{}
Upon measurement, a qubit loses its quantumness and its superposition state (\ref{eq:qbitState}) is transformed into classical information, either $\ket{0}$ with the probability $ \abs{\phi_0}^2$ or $\ket{1}$ with the probability $ \abs{\phi_1}^2$. 

A system consisting of $K$ qubits is a $2^K$-state quantum-mechanical system whose state can be represented by any normalised linear combination of $2^K$ basis states following
\begin{eqnarray}\label{eq:multipleQubits}
	\ket{\phi} &=& \phi_0 \ket{0}\otimes\ldots\otimes\ket{0}\otimes\ket{0}+ \phi_1\ket{0}\otimes\ldots\otimes\ket{0}\otimes \ket{1}  + \phi_{2^K-1} \ket{1}\otimes\ldots\otimes\ket{1}\otimes\ket{1}  \,,
\end{eqnarray}
where $\otimes$ denotes the Kronecker product\footnote{The Kronecker product $\otimes$ is commonly used in quantum computing. The Kronecker product of two matrices $\vb{A} \in \mathbb{C}^{m\times n}$ and $\vb{B} \in \mathbb{C}^{p\times q}$ results in a matrix $\vb{C}\in \mathbb{C}^{mp\times nq}$ as $C_{ab} = A_{ij}B_{kl}$, where $a=pi+k$ and $b=qj+l$ with $i=0,\ldots,m-1$, $j=0,\ldots,n-1$, $k=0,\ldots,p-1$, and $l=0,\ldots,q-1$.\removedetails{ More explicitly, $\vb{C}$ is expressed as the following block matrix
	\begin{eqnarray}
		\vb{C}=\mqty[A_{00}\vb{B}&\ldots&A_{00}\vb{B}\\ \vdots&\ddots&\vdots \\A_{{m-1}\,0}\vb{B}&\ldots&A_{{m-1}\,{n-1}}\vb{B}]\,.
\end{eqnarray}}{}}, and $\phi_0$, $\ldots$, $\phi_{2^K-1}$ are complex numbers satisfying the normalisation condition $\sum_{i=0}^{2^{K}-1} \abs{\phi_i}^2 = 1$\removedetails{, and $\ket{b_0} \otimes \ldots \otimes\ket{b_{K-1}}$ for $b_i\in \{0, 1\}\,,\; \forall i$, are the orthonormal bases of a $2^K$-dimensional complex Hilbert space. For simplicity, $\ket{b_0} \otimes \ldots \otimes\ket{b_{K-1}}$ is usually written as a binary string $ \ket{b_0 \ldots b_{K-1}}$ in the literature.}{.}

\subsection{Principle of quantum annealing}
The objective of the quantum annealing is to find the ground state of the Hamiltonian $\vb{H}$, \emph{i.e.} solving
\begin{eqnarray}
	\ket{\phi_0} = \argmin{\ket{\phi}} \bra{\phi}\vb{H}\ket{\phi}\,,
\end{eqnarray} 
where the bra notation $\bra{\bullet}$ corresponds to the conjugate transpose of the ket notation $\ket{\bullet}$. The state $\ket{\phi_0}$ is called the ground state of $\vb{H}$ and $E_0 =\bra{\phi_0}\vb{H}\ket{\phi_0}$ is the ground-state energy which corresponds to the minimum eigenvalue of $\vb{H}$ with $\ket{\phi_0}$ being its associated eigenvector \cite{tilly2022variational}.

Quantum annealing is based on the principle of adiabatic quantum computation. The basic idea is to start with a simple Hamiltonian $\vb{H}_0$ whose ground state is known and which can be prepared easily, and then gradually transform this Hamiltonian into the problem Hamiltonian $\vb{H}$ whose ground state encodes the solution to the optimisation problem at hand. The time-dependent Hamiltonian of the system reads
\begin{eqnarray}\label{eq:HQA}
	\vb{H}_\text{QA}\left(t\right) = A(t) \vb{H}_0 + B(t)\vb{H}\,,
\end{eqnarray}
where $A(t)$ and $B(t)$ are annealing functions satisfying $A(0)=1 \gg B(0)$ and $B(t_{\text{a}})=1\gg A(t_{\text{a}})$ where $t_{\text{a}}$ is the annealing time. By slowly varying the Hamiltonian from 0 to $t_{\text{a}}$, the system remains in its ground state throughout the process as the result of the adiabatic theorem \cite{born1928beweis}, and, eventually, the final state of the system corresponds to the sought solution of the considered optimisation problem. The details of such a annealing process can be found in the review \cite{albash2018adiabatic}.

\subsection{Ground state of an Ising Hamiltonian} \label{subsec:Ising}

We consider the problem of finding the ground state of the Ising Hamiltonian \cite{kadowaki1998quantum,brooke1999quantum,tanaka2017quantum}, which is represented by an undirected graph $(V, E)$ where $V=\{0, \ldots, K-1\}$ is a set of $K$ qubits and $E$ specifies the set of interactions between two qubits, \emph{i.e.} $E \subset \{(i,j) |i\in V, j \in V, \text{ and } i < j\}$ as
\begin{eqnarray}\label{eq:IsingHamilton}
	\vb{H} = \sum_{i\in V} h_i \vb{H}_i + \sum_{(i,j)\in E}J_{ij}\vb{H}_{ij} \,,
\end{eqnarray}
where $h_i \neq 0 $ and $J_{ij}\neq 0$, with $i\in V$ and $(i,j)\in E$, are constants, and $ \vb{H}_i$, and $\vb{H}_{ij}$ are the Pauli-Z operators respectively applied on the qubit $i$ and on the qubits $i$ and $j$ as
\begin{eqnarray}
	\vb{H}_i&=& \underbrace{\vb{I}}_{{0}}\otimes\ldots\otimes  \vb{I}\otimes\underbrace{\vb{Z}}_{{i}}\otimes \vb{I} \otimes \ldots \otimes \underbrace{\vb{I}}_{K-1}\,,\text{ and } \\
		\vb{H}_{ij}&=& \underbrace{\vb{I}}_{0}\otimes\ldots\otimes \vb{I}\otimes \underbrace{\vb{Z}}_i\otimes \vb{I} \otimes\ldots\otimes \vb{I}\otimes\underbrace{\vb{Z}}_j\otimes \vb{I} \otimes \ldots \otimes \underbrace{\vb{I}}_{K-1}\,, \\
		&&\text{with } \vb{Z} = \mqty[1&0\\0&-1] \text{ and } \vb{I}= \mqty[1&0\\0&1]\,.
\end{eqnarray} 

It can be seen that $\vb{H}$ in Eq. (\ref{eq:IsingHamilton}) is a $2^K\times 2^K$ diagonal operator in the computational basis. Consequently, its ground state corresponds to the one of its computational basis. Assuming $\ket{\phi}=\ket{b_0\ldots b_{K-1}}$, with $b_i\in \{0,1\} \,\; \forall i=0,\,\ldots,\,K-1$, is the computational basis of $\vb{H}$, its associated eigenvalue reads\footnote{This relation results from the definition of Eq. (\ref{eq:IsingHamilton}) and the use of the following properties of $\vb{Z}$, $\vb{H}_i$, and $\vb{H}_{ij}$:
\begin{eqnarray}
	\vb{Z}\ket{b_i}&=&(-1)^{b_i}\ket{b_i}\,,\nonumber \\
	\vb{H}_i \ket{\phi} &=& \ket{b_0 \ldots b_{i-1}} \otimes \vb{Z}\ket{b_i} \otimes \ket{b_{i+1}\ldots b_{K-1}} = (-1)^{b_i} \ket{\phi} \nonumber \,,\text{ and}\nonumber \\
	\vb{H}_{ij} \ket{\phi} &=&   \ket{b_0 \ldots b_{i-1}} \otimes \vb{Z}\ket{b_i} \otimes \ldots \otimes b_{j-1} \otimes \vb{Z}\ket{b_j} \otimes  \ket{b_{j+1}\ldots b_{K-1}} = (-1)^{b_i} (-1)^{b_j} \ket{\phi} \nonumber \,.
\end{eqnarray}
}
\begin{eqnarray}\label{eq:minSpin}
\mathcal{F}_\text{Ising} = \sum_{i\in V} h_i q_i + \sum_{(i,j)\in E} J_{ij} q_i q_j = \vb{q}^T\vb{h} + \vb{q}^T\vb{J}\vb{q}\,,
\end{eqnarray}
where $q_i=(-1)^{b_i}\in  \{-1,1\}$, which is known as the spin variable, $\vb{q}=[q_i \; \forall i\in V]$ is the spin vector, $\vb{h}=[h_i \;\forall i\in V]$ is the biases vector, and $\vb{J}=[J_{ij} \text{ if } (i,j)\in E; 0 \text{ otherwise}]$ is a strictly upper diagonal matrix.

As a result, finding the ground state of the Ising Hamiltonian (\ref{eq:IsingHamilton}) is equivalent to minimising the quadratic function of the spin variables (\ref{eq:minSpin}), yielding the optimisation problem
\begin{eqnarray}
	\vb{q} = \argmin{\vb{q}^{\prime}} \mathcal{F}_\text{Ising}\left(\vb{q}^{\prime};\, \vb{h},\, \vb{J}\right) \label{eq:QAdefSpin}\,,
\end{eqnarray}
where $\vb{h}$ and $\vb{J}$ are the user programmable parameters.

The minimisation of $\mathcal{F}_\text{Ising}$ stated by Eq. (\ref{eq:QAdefSpin}) belongs to the class of problems known as quadratic unconstrained binary optimisation (QUBO). Indeed, using the spin-binary variable transformation $q_i=2b_i-1: \{0,1\}\rightarrow \{-1,1\}\,,$ and the property $b_i^2=b_i\;\forall i$, the minimisation of $\mathcal{F}_\text{Ising}$ stated by Eq. (\ref{eq:QAdefSpin}) is equivalent to the minimisation of $\mathcal{F}_\text{QUBO}$ which takes the form
\begin{eqnarray}\label{eq:QAdef}
 \vb{b}= \argmin{\vb{b}^{\prime}}{\mathcal{F}_\text{QUBO}}\left(\vb{b}^{\prime};\, \vb{A}\right) \text{  with  }	\mathcal{F}_\text{QUBO}\left(\vb{b};\, \vb{A}\right) = \sum_{(i,j) \in E \cup \{(i,i) \,\forall i\in V\} } A_{ij} b_i b_j = \vb{b}^T \vb{A}\vb{b}\,,
\end{eqnarray}
where $\vb{A}=[A_{ij}\text{ for } (i,j) \in E  \cup \{(i,i) \,\forall i\in V\}; 0 \text{ otherwise} ]$ is the QUBO matrix, a user programmable parameter, and $\vb{b}=\mqty[b_0&\ldots&b_{K-1}]^T$ is the vector of the unknown binary variables.

\subsection{Ising Hamiltonian-based quantum annealing}

Quantum annealing considering the Ising Hamiltonian in Eq. (\ref{eq:HQA}) can be implemented on a hardware called quantum annealer \cite{johnson2011quantum}, which, from the computational point-of view, provides a black-box optimiser for solving the minimisation problems (\ref{eq:QAdefSpin}) or (\ref{eq:QAdef}) defined by their user programmable parameters, respectively $\vb{h}$ and $\vb{J}$, or $\vb{A}$. Different problems can be mapped to such a form through these programmable parameters  $\vb{h}$ and $\vb{J}$, or $\vb{A}$.
}

\secondreviewer{Quantum annealers are devices designed to solve optimisation problems using quantum mechanics principles. Their outputs are generally probabilistic rather than deterministic since quantum mechanics inherently introduces probabilities into the outcomes. Quantum annealers may not converge to the global minimum on a single run due to various factors, such as environmental noises, hardware imperfections, and pre- and post-processing errors. In practice, the annealing cycle needs to be repeated multiple times per input to obtain multiple candidate solutions and the cycle resulting into the smallest energy is chosen to provide the solution to the optimisation problem \cite{yarkoni2022quantum}. The probability of finding the optimal or near-optimal solution is generally higher when increasing the number of annealing cycles per input. In general, the following steps are followed to solve a problem with quantum annealers (see \emph{e.g.} \cite{MCGEOCH2020169}):
	\begin{enumerate}
		\item  The problem under consideration ($P_1$) has to be transformed into an Ising or a QUBO form ($P_2$) following, respectively Eq. (\ref{eq:QAdefSpin}) or Eq. (\ref{eq:QAdef});
		\item  An embedding technique is applied to transform $P_2$ into $P_3$, which matches the quantum processing units (QPU) topology representing the physical connectivity between the qubits;
		\item $P_3$ is sent to the QPU together with other user parameters. The graph of $P_3$ is limited by the number of qubits and their available interactions in the QPU. The solution of $P_3$ from the QPU is returned as a set of samples, in which the best solution is selected as the case corresponding to the smallest energy. 
\end{enumerate}}

\firstreviewer{In recent years, the progress in developing quantum annealing hardware has gained remarkable acceleration, in particular available D-Wave QPUs are now being used in various applications \cite{yarkoni2022quantum,rajak2023quantum}. This work considers the most recent D-Wave QPU (Advantage QPU) possessing more than 5000 qubits with more than 35000 couplers. The maximum size of a fully connected Ising graph that can be embedded in the QPU is equal to 177 since each qubit is only connected to a reduced number of other qubits \cite{DwaveReport}. A sparse graph with 1 to the maximum number of available qubits can be solved as long as it fits the QPU topology.}

\secondreviewer{The solution time offered by the D-Wave quantum annealers depends on the complexity of the optimisation problem, the number of qubits involved, and the other settings. It can be measured by the QPU access time, which mainly consists of the programming time (with a magnitude of the order of the millisecond) and the sampling time per annealing cycle (including annealing time, readout time, and delay time with a magnitude of the order of the microsecond), and by the number of cycles (from 1 to 10000 cycles in Advantage QPU). With the current Advantage QPU, the maximum time that a submitted problem is allowed to run is 1 second. More details on the solution time on D-Wave QPU can be found in \cite{willsch2022benchmarking}.}

\section{Variational formulation of the history-dependent mechanical boundary value problem} \label{sec:mechaBVP}

In this section, the mechanical boundary value problem is formulated as a minimisation problem using the variational updates formulation \cite{ortiz1999variational,radovitzky1999error,weinberg2006variational}. However, contrarily to classical finite element approaches for which the material model is solved locally whilst the finite element degrees of freedom are solved at the structural scale in a nested way, we develop a variational formulation at the structural scale.
This approach is motivated by the fact that these minimisation problems can be performed in the quantum annealers after a suitable continuous-discrete variables transformation. The resulting hybrid framework combining classical computers and quantum annealers will be detailed in Section \ref{sec:QA-SQP}.
It is thus advantageous to solve both the material model and the structural governing equations using annealing whilst minimising the interface steps between the different hardware.

\subsection{Governing equations}
Considering a body $V\in \mathbb{R}^d$ where $d$ is the problem dimension. Its boundary $\partial V$ is divided into two non-overlapping parts: $\partial_D V$, the Dirichlet boundary where a prescribed displacement $\vb{u}_0$ is enforced, and $\partial_N V$, the Neumann boundary where the traction per unit reference surface $\vb{t}_0$ is prescribed. The quasi-static equilibrium equations over the body $V$ are given as  
\begin{eqnarray}
	&&\grad\cdot \bm{\sigma}\left(\vb{x}\right)+ \vb{b}_0\left(\vb{x}\right)=\vb{0} \,\,\forall \vb{x}\in V \label{eq:balance1} \,,\\
	&&\vb{u}\left(\vb{x}\right) = \vb{u}_0\left(\vb{x}\right) \,\,\forall \vb{x}\in  \partial_D V \,, \text{ and } \label{eq:balance2}\\
	&&\bm{\sigma}\left(\vb{x}\right) \cdot\vb{n}\left(\vb{x}\right)  = \vb{t}_0\left(\vb{x}\right) \,\,\forall\vb{x}\in \partial_N V \label{eq:balance3}\,,
\end{eqnarray}
where $\grad$ is the gradient operator, $\bm{\sigma}$ is the Cauchy stress tensor, $\vb{b}_0$ is the volumetric force, and $\vb{n}$ is the outward unit normal to $\partial_N V$. By assuming a small strain formulation, a constitutive relation must be introduced in order to solve the displacement field $\vb{u}$ as
\begin{eqnarray} \label{eq:constitutiveLaw1}
	\begin{cases}
		&\bm{\sigma}\left(\vb{x}\right)  = \bm{\sigma}\left( \bm{\varepsilon}\left(\vb{x}\right) ; \vb{q}\left(\vb{x}\right) \right) \,, \text{ and } \\
		& \text{evolution laws for } \vb{q} \,,
	\end{cases}
\end{eqnarray}
where $\bm{\varepsilon}=\grad \otimes^s{\vb{u}}$ is the deformation tensor with $\varepsilon_{ij}=\cfrac{1}{2}\left(\cfrac{\partial u_i}{\partial x_j} + \cfrac{\partial u_j}{\partial x_i}\right)$, and $\vb{q}$ is the vector of all the internal variables used to describe a history-dependent stress-strain relationship. In this work, we follow the variational principle in which the mechanical boundary problem can be described as a double-minimisation problem: one with respect to the displacement field $\vb{u}\left(\vb{x}\right)\,,\;\forall \vb{x}\in V$, and one with respect to the internal variables field $\vb{q}\left(\vb{x}\right)\,,\;\forall \vb{x}\in V$.

\subsection{Double-minimisation formulation}\label{subsec:doubleMinimisationProblem}

Quantum annealing is particularly efficient to solve a minimisation problem, which is fully compatible with the underlying physics of the numerical methods developed in computational mechanics. Indeed, not only the mechanical balance equation (\ref{eq:balance1}-\ref{eq:balance3}), but also the constitutive model (\ref{eq:constitutiveLaw1}) can be cast under the form of an functional minimisation.
For this purpose, the minimisation of energy functions is first formulated in an incremental way at each loading increment and then solved, leading to the solution in terms of not only the displacement field but also of the internal variables field. To this end, the variational updates formulation pioneered for the visco-plastic constitutive models \cite{ortiz1999variational,radovitzky1999error}, and later extended to porous metal plasticity \cite{weinberg2006variational}, elasto-plastic solids \cite{mosler2010implementation}, thermo-mechanical coupling \cite{stainier2010study, yang2006variational}, visco-elastic solids \cite{fancello2008variational,vassoler2012variational}, elasto-plastic-damage solids \cite{kintzel2010coupled,kintzel2011incremental}, \secondreviewer{and multi-scale simulations \cite{miehe2002strain,brassart2011variational}}, is reframed in order to express the finite element and constitutive model resolutions as a double-minimisation problem expressed at the global or structural scale.

\secondreviewer{
The existence of a Helmholtz free energy $\Psi$ is postulated under the following form
\begin{eqnarray}\label{eq:psi}
	\Psi = \Psi\left(\bm{\varepsilon}, \vb{q}\right)\,,
\end{eqnarray}
explicitly depending on the strain tensor $\bm{\varepsilon}$ and on a collection of history-dependent variables $\vb{q}$ introduced to model history dependency. The Clausius-Duhem inequality imposes the following restriction
\begin{eqnarray}
	\mathcal{D} = \bm{\sigma}:\dot{\bm{\varepsilon}} - \dot{\Psi} \geq 0\,,
\end{eqnarray}
where $\mathcal{D}$ is the dissipation.
This inequality becoming an equality for any reversible process, Eq. (\ref{eq:psi}) results in the following stress-strain relationship
\begin{eqnarray}\label{eq:stressDef}
	\bm{\sigma} = \frac{\partial \Psi}{\partial \bm{\varepsilon}}\,,
\end{eqnarray}
whilst the non-negativeness constraint on the dissipation for any reversible or irreversible process implies
\begin{eqnarray}\label{eq:conjugateforce}
	\mathcal{D}\left(\vb{Y},\dot{\vb{q}}\right) = \vb{Y}\cdot \dot{\vb{q}} \geq 0 \text{ with } \vb{Y}=-\cfrac{\partial \Psi}{\partial \vb{q}} \,,
\end{eqnarray}
where $\vb{Y}$ is the vector of internal forces conjugate to $\vb{q}$, and the operator ``$\cdot$'' denotes the inner product between two vectors.

For problems involving history dependency, \emph{i.e.} $\vb{q} \neq \emptyset$, the evolution laws specifying the evolution of the internal variables $\vb{q}$ must be provided. In the case of standard materials \cite{halphen1975materiaux}, it is assumed that there exists a dissipation pseudo-potential $\Theta\left(\dot{\vb{q}}\right)$ and its convex dual $\Theta^\star\left(\vb{Y}\right)$ derived by considering the Legendre transformation and by postulating the maximum dissipation, yielding
\begin{eqnarray}\label{eq:dissipation}
	\Theta\left(\dot{\vb{q}}\right) = \max_{\vb{Y}} \left[ \vb{Y}\cdot \dot{\vb{q}} - \Theta^\star\left(\vb{Y}\right) \right]\,.
\end{eqnarray}
The evolution law for $\dot{\vb{q}}$ thus reads
\begin{eqnarray}\label{eq:dissipationEvolution}
	\dot{\vb{q}} = \frac{\partial  \Theta^\star\left(\vb{Y}\right)}{\partial \vb{Y}}, \text{ or equivalently } \vb{Y} =\frac{\partial  \Theta\left(\dot{\vb{q}}\right)}{\partial \dot{\vb{q}}} \,.
\end{eqnarray}

Using Eq. (\ref{eq:conjugateforce}), the local power functional $\mathcal{E}\left(\dot{\varepsilon},\,\dot{\vb{q}}\right)$ is defined from new independent variables $\left(\dot{\bm{\varepsilon}},\,\dot{\vb{q}}\right)$ as 
\begin{eqnarray}\label{eq:EnergyFunc}
  \mathcal{E}\left(\dot{\bm{\varepsilon}},\,\dot{\vb{q}}\right) = \dot{\Psi}+ \Theta =\frac{\partial \Psi}{\partial \bm{\varepsilon}}:\dot{\bm{\varepsilon}}-\vb{Y}\cdot \dot{\vb{q}}+\Theta \,.
\end{eqnarray}
It is clear that this local power functional $\mathcal{E}\left(\dot{\varepsilon},\,\dot{\vb{q}}\right)$ has to be minimised with respect to the internal variable rate since, using Eq. (\ref{eq:dissipationEvolution}), one has
\begin{eqnarray}\label{eq:EnergyFuncRate}
 \frac{\partial \mathcal{E}\left(\dot{\bm{\varepsilon}},\,\dot{\vb{q}}\right)}{\partial \dot{\vb{q}}}  = -\vb{Y}+\frac{\partial  \Theta\left(\dot{\vb{q}}\right)}{\partial \dot{\vb{q}}}=\vb{0} \,.
\end{eqnarray}
Following \cite{ortiz1999variational}, the effective power functional $\mathcal{E}^{\text{eff}}\left(\dot{\bm{\varepsilon}}\right)$ is thus defined by
\begin{eqnarray}\label{eq:EffEnergyFuncRate}
\mathcal{E}^{\text{eff}}\left(\dot{\bm{\varepsilon}}\right)=\min_{\dot{\vb{q}}} \mathcal{E}\left(\dot{\bm{\varepsilon}},\,\dot{\vb{q}}\right)\;\text{  with  } \; \frac{\partial \mathcal{E}^{\text{eff}}\left(\dot{\bm{\varepsilon}}\right)}{\partial \dot{\bm{\varepsilon}}}  =\frac{\partial \Psi}{\partial \bm{\varepsilon}} =\bm{\sigma} \,.
\end{eqnarray}

Considering now an arbitrary time step $[t_n,\,t_{n+1}]$, the incremental form \cite{ortiz1999variational} of Eq. (\ref{eq:EnergyFuncRate}) reads
\begin{eqnarray}\label{eq:J2EnergyFunc}
	\Delta\mathcal{E}_{n+1}\left({\bm{\varepsilon}}_{n+1},\,\vb{q}_{n+1};\;{\bm{\varepsilon}}_{n},\,\vb{q}_{n} \right)  =  \int_{t_n}^{t_{n+1}} \mathcal{E} dt\;,
\end{eqnarray}
where $\Delta \bullet_{n+1} = \bullet_{n+1} - \bullet_{n}$ denotes the increment of an arbitrary quantity $\bullet$, and $\bullet_{n+1}$ and $\bullet_{n}$ are respectively its values at configurations $t_{n+1}$ and $t_{n}$.
For a strain-driven constitutive model, $\bm{\varepsilon}_{n+1}$ and $\vb{q_n}$ are known and $\vb{q}_{n+1}$ is obtained from the incremental form of Eq. (\ref{eq:EffEnergyFuncRate}), yielding
\begin{eqnarray}\label{eq:internalVartionalLocal}
\vb{q}_{n+1} = \argmin{ \vb{q}^{\prime}_{n+1}} \Delta\mathcal{E}_{n+1}\left( {\bm{\varepsilon}}_{n+1},\, {\vb{q}}^{\prime}_{n+1}\right)\;  \text{  and  } \; \Delta\mathcal{E}_{n+1}^\text{eff}\left({\bm{\varepsilon}}_{n+1}\right)= \min_{\vb{q}_{n+1}} \Delta\mathcal{E}_{n+1}\left({\bm{\varepsilon}}_{n+1}, \, {\vb{q}}_{n+1}\right) \,,
\end{eqnarray}
where from now on the known terms at configuration $n$ are omitted in the functional argument for conciseness. Eventually, the stress tensor arises from
\begin{eqnarray}\label{eq:internalVartionalLocal2}
\bm{\sigma}_{n+1} = \frac{\partial \Delta\mathcal{E}^{\text{eff}}_{n+1}\left({\bm{\varepsilon}}_{n+1}\right)}{\partial {\bm{\varepsilon}}_{n+1}}\,. 
\end{eqnarray}

One can thus define the power functional over the domain volume $V$ as
\begin{eqnarray}\label{eq:virtualPower2}
	\dot{\Phi} \left( \dot{\vb{u}},\, \dot{\vb{q}}\right) = \int_{V} \mathcal{E}\left( {\bm{\varepsilon}}(\dot{\vb{u}}),\, \dot{\vb{q}}\right) \,dV - \dot{W}^\text{ext}(\dot{\vb{u}}) \,,
\end{eqnarray}
where $\dot{W}^\text{ext}$ is the external power which is given as
\begin{eqnarray}
	\dot{W}^\text{ext} = \int_V \vb{b}_0\cdot\dot{\vb{u}}\,dV + \int_{\partial_N V} \vb{t}_0 \cdot\dot{\vb{u}}\,dA\,.
\end{eqnarray}
Considering the time step $[t_n,\,t_{n+1}]$ and assuming $\vb{u}_n\left(\vb{x}\right) $ and $\vb{q}_n\left(\vb{x}\right)\,,\; \forall \vb{x}\in V$, are known, Eq. (\ref{eq:virtualPower2}) can be integrated over $[t_n,\,t_{n+1}]$, leading to
\begin{eqnarray}\label{eq:energyVar}
	\Delta \Phi_{n+1}\left(\vb{u}_{n+1},\, \vb{q}_{n+1}\right) = \int_{t_n}^{t_{n+1}}\dot{\Phi} \,dt = \int_{V} \Delta \mathcal{E}_{n+1}\left(\bm{\varepsilon}\left(\vb{u}_{n+1}\right),\, \vb{q}_{n+1}\right)\,dV - \Delta W^\text{ext}_{n+1}\left(\vb{u}_{n+1}\right)\,.
\end{eqnarray}
For a given value of $\vb{u}_{n+1}$, the minimisation problems (\ref{eq:internalVartionalLocal}) can be performed locally to find $\vb{q}_{n+1}\left(\vb{x}\right)\,,\;\forall\vb{x}\in V$, leading to the overall optimisation problem 
\begin{eqnarray}
	\vb{q}_{n+1} = \argmin{{\vb{q}^{\prime}_{n+1}}} \Delta\Phi_{n+1}\left({\vb{u}}_{n+1}, {\vb{q}^{\prime}_{n+1}}\right) \label{eq:internalVartional} \,.
\end{eqnarray}
Combining Eqs. (\ref{eq:energyVar}, \ref{eq:internalVartional}) leads to
\begin{eqnarray}
	\Delta \Phi_{n+1}^\text{eff}\left({\vb{u}_{n+1}}\right) =  \int_{V} \Delta \mathcal{E}_{n+1}^\text{eff}\left({\bm{\varepsilon}}\left(\vb{u}_{n+1}\right)\right)\,dV - \Delta{W}_{n+1}^\text{ext}\left(\vb{u}_{n+1}\right)\,,
\end{eqnarray}
where $ \mathcal{E}_{n+1}^\text{eff}$ is estimated from the solution of the optimisation problem (\ref{eq:internalVartional}). The updated displacement field ${\vb{u}}_{n+1}$ finally follows from the minimisation problem
\begin{eqnarray}\label{eq:dispMin}
	\begin{cases}
		&{\vb{u}}_{n+1} = \argmin{{\vb{u}}^{\prime}_{n+1}} \Delta \Phi_{n+1}^\text{eff}\left({\vb{u}^{\prime}_{n+1}}\right) \,,\\
		&\text{ with } {\vb{u}}^{\prime}_{n+1} \text{ kinematically admissible.}
	\end{cases}
\end{eqnarray}

These expressions will be particularised to the considered material models in Section \ref{sec:constitutiveModels}.
}

The double-minimisation algorithm to be solved, \emph{i.e.} the minimisation problem (\ref{eq:internalVartional}) and the minimisation problem (\ref{eq:dispMin}), is illustrated in Alg. \ref{alg:DoubleMinimisation}, in which the solution at time $t_n$ is known and the problem is solved at time $t_{n+1}$.

\begin{algorithm}[htb!]
	\caption{Double-Minimisation Algorithm.}
	\label{alg:DoubleMinimisation}
	\begin{algorithmic}[1]
		\Require $\vb{u}_{n}$, $\vb{q}_n$;
		\Ensure $\vb{u}_{n+1}$, $\vb{q}_{n+1}$;
		\State initialise ${\vb{u}}_{n+1}\gets\vb{u}_{n}$ and $\vb{q}_{n+1}\gets\vb{q}_n$;
		\State apply boundary conditions;
		\Repeat
		\State $\vb{u}_{n+1}\gets$ solve the problem (\ref{eq:dispMin});
		\State $\vb{q}_{n+1}\gets$ solve the problem (\ref{eq:internalVartional});
		\Until{convergence}
	\end{algorithmic}
\end{algorithm}

\subsection{Constitutive models}\label{sec:constitutiveModels}

\secondreviewer{In this section, the variational formulation is detailed for two usual material models: linear elasticity and $J_2$-von Mises elasto-plasticity.}

\subsubsection{Linear elasticity}
In the case of a linear elastic material, one has $\vb{q} = \emptyset$ and the free energy reads
\secondreviewer{
	\begin{eqnarray}\label{eq:psiEl}
		\Psi &= & \cfrac{K}{2}\left[\trace{\bm{\varepsilon}}\right]^2 + \mu \dev{\bm{\varepsilon}}:\dev{\bm{\varepsilon}}\,,
	\end{eqnarray}
	where $K$ and $\mu$ are the bulk and shear moduli, respectively.} The stress tensor $\bm{\sigma}=\cfrac{\partial  \Psi}{\partial \bm{\varepsilon}}$ and the tangent operator $\vb{C} = \cfrac{\partial \bm{\sigma}}{\partial\bm{\varepsilon}}$ follow directly from Eq. (\ref{eq:psiEl}).
Since there is no internal variable, the minimisation problem (\ref{eq:internalVartional}) is trivial and the minimisation problem (\ref{eq:dispMin}) is solved by considering directly $\Phi\left(\vb{u}\right)$ from Eq. (\ref{eq:energyVar}), which is rewritten as
\begin{eqnarray}
	\Phi\left(\vb{u}\right)=\int_{V} \Psi\left(\vb{u}\right) \,dV - {W}^\text{ext}\,.
\end{eqnarray}

\subsubsection{$J_2$-elasto-plasticity} \label{subsec:J2}

\secondreviewer{
In this model, the total strain tensor $\bm{\varepsilon}$ is decomposed into a reversible elastic part $\bm{\varepsilon}^e$ and an irreversible plastic part $\bm{\varepsilon}^{\text{p}}$, such that $\bm{\varepsilon} = \bm{\varepsilon}^{\text{e}} + \bm{\varepsilon}^{\text{p}}$. The collection of internal variables reads 
\begin{eqnarray}
	\vb{q}=\{\bm{\varepsilon}^{\text{p}},\,\gamma \}\,,
\end{eqnarray}
where $\gamma$ is a scalar characterising the isotropic hardening evolution and which is related to the plastic strain tensor $\bm{\varepsilon}^{\text{p}}$ through 
\begin{eqnarray}\label{eq:plasticFlow}
	\dot{\bm{\varepsilon}}^{\text{p}} = \dot{\gamma}\vb{N}\,,
\end{eqnarray}
where $ \dot{\gamma}\geq 0$ and $\vb{N}$ is the plastic normal subjected to the constraints
\begin{eqnarray}\label{eq:plasticConstraints}
	\trace{\vb{N}} = 0 \text{ and } \vb{N}:\vb{N} = \frac{3}{2}\,.
\end{eqnarray}

The free energy and dissipation pseudo-potential are chosen as follows
\begin{eqnarray}
	\Psi &= & \cfrac{K}{2}\left[\trace{\bm{\varepsilon}}\right]^2 + \mu \left[\dev{\bm{\varepsilon}} -\bm{\varepsilon}^\text{p} \right]:\left[\dev{\bm{\varepsilon}} -\bm{\varepsilon}^\text{p} \right] \,,  \text{ and}\\
	\Theta &=& \begin{cases}
		\left[\sigma_y^0+ R\left(\gamma\right)\right] \dot{\gamma} & \text{ if } \dot{\gamma} \geq 0\,,\\
		\infty & \text{ otherwise}\,,
	\end{cases}
\end{eqnarray}
where $K$ and $\mu$ are respectively the bulk and shear moduli, $\sigma_y^0$ is the initial yield stress characterising the onset of plasticity and $R\left(\gamma\right)$ is the hardening law introduced to capture the isotropic hardening effect. The corresponding dual potential is
\begin{eqnarray}
	\Theta^\star = \begin{cases}
		0 &\text{ if } \sigma_{\text{eq}}-\sigma_y^0 - R \leq 0\,, \\
		\infty& \text{ otherwise}\,.
	\end{cases}\,,
\end{eqnarray}
where $\sigma_{\text{eq}}$ is the equivalent von Mises stress of the stress tensor $\bm{\sigma}$.\footnote{\secondreviewer{There are other possibilities to define the free energy and dissipation pseudo-potential, see \emph{e.g.} the works \cite{ortiz1999variational,miehe2002strain,brassart2011variational} in which the free energy consists of the reversible and irreversible parts. As an example, in  \cite{ortiz1999variational}, the free energy functional was defined by
		\begin{eqnarray}
			\Psi = \cfrac{K}{2}\left[\trace{\bm{\varepsilon}}\right]^2 + \mu \left[\dev{\bm{\varepsilon}} -\bm{\varepsilon}^\text{p} \right]:\left[\dev{\bm{\varepsilon}} -\bm{\varepsilon}^\text{p} \right] + \int_{0}^{\gamma} \left[\sigma_y^0 
			+ R\left(\gamma^\prime \right)\right]\,d\gamma^\prime\,,
		\end{eqnarray}
		and the dissipation pseudo-potential was defined as an indicator specifying the admissible domain as
		\begin{eqnarray}
			\Theta &=& \begin{cases}
				0& \text{ if } \dot{\gamma} \geq 0\,,\\
				\infty & \text{ otherwise}\,.
			\end{cases}
		\end{eqnarray}
		Although different definitions can be employed, the same functional to be minimised is eventually obtained.
}}

Whilst the stress tensor $\bm{\sigma}$ follows from Eq. (\ref{eq:stressDef}) or Eq. (\ref{eq:internalVartionalLocal2}), this requires to know the internal state of the system defined by $\{\bm{\varepsilon}^{\text{p}}\,,\gamma\}$.
Considering a time step $[t_n, t_{n+1}]$, all quantities at time $t_n$ are known.
With the introduction of the plastic flow direction $\vb{N}$ (\ref{eq:plasticFlow}), one has
\begin{eqnarray}
	\Delta{\bm{\varepsilon}}_{n+1}^{\text{p}} = {\bm{\varepsilon}}_{n+1}^{\text{p}}-{\bm{\varepsilon}}_{n}^{\text{p}} = \Delta{\gamma}_{n+1} \vb{N}_{n+1}\,,
\end{eqnarray} 
and the internal state of the model can thus be defined by
\begin{eqnarray}
	{\vb{q}}_{n+1} =\{\bm{\varepsilon}_{n+1}^{\text{p}},\,\gamma_{n+1}\}\equiv\{\gamma_{n+1},\,\vb{N}_{n+1}\} \,.
\end{eqnarray}
The energy increment of the energy functional (\ref{eq:J2EnergyFunc}) thus reads
\begin{eqnarray} \label{eq:energyfunctionJ2}
	&&  \Delta \mathcal{E}_{n+1}\left(\vb{u}_{n+1},\,\Delta{\gamma}_{n+1},\,\vb{N}_{n+1}\right) = \nonumber \\
	&& \cfrac{K}{2}\left[\trace{\bm{\varepsilon}_{n+1}}\right]^2 + \mu \left[\dev{\bm{\varepsilon}_{n+1}} -\bm{\varepsilon}^\text{p}_{n}-\Delta\gamma_{n+1}\vb{N}_{n+1} \right]:\left[\dev{\bm{\varepsilon}_{n+1}} -\bm{\varepsilon}^\text{p}_{n}-\Delta\gamma_{n+1}\vb{N}_{n+1} \right]\nonumber \\
	&&-\cfrac{K}{2}\left[\trace{\bm{\varepsilon}_{n}}\right]^2 + \mu \left[\dev{\bm{\varepsilon}_{n}} -\bm{\varepsilon}^\text{p}_{n} \right]:\left[\dev{\bm{\varepsilon}_{n}} -\bm{\varepsilon}^\text{p}_{n} \right] + \int_{\gamma_n}^{\gamma_n+\Delta \gamma_{n+1}} \left[\sigma_y^0 
	+ R\left(\gamma^\prime \right)\right]\,d\gamma^\prime \label{eq:energyJ2Inc_body} \,.
\end{eqnarray}
Therefore, at a material point $\vb{x}$, the internal variables arise from the minimisation of the convex energy function (\ref{eq:energyfunctionJ2}) under the constraints (\ref{eq:plasticConstraints}), yielding
\begin{eqnarray}
	\{\Delta{\gamma}_{n+1}\left(\vb{x}\right),\, \vb{N}_{n+1}\left(\vb{x}\right)\} = \argmin{\{\Delta{\gamma}^{\prime}_{n+1},\, \vb{N}^{\prime}_{n+1}\}} \Delta \mathcal{E}_{n+1}\left(\vb{u}_{n+1},\,\Delta{\gamma}^{\prime}_{n+1},\,\vb{N}^{\prime}_{n+1}\right) \,, \label{eq:variationalJ2}  \\
	\text{satisfying } \Delta{\gamma}_{n+1} \geq\, 0,\; \trace{\vb{N}_{n+1}} = 0\,,\; \text{ and } \vb{N}_{n+1}:\vb{N}_{n+1} = \cfrac{3}{2} \label{eq:variationalJ2Constraints}\,.
\end{eqnarray}
The constrained minimisation problem (\ref{eq:variationalJ2}, \ref{eq:variationalJ2Constraints}) yields the well-known radial return scheme in the conventional $J_2$-plasticity model \cite{de2011computational, brassart2011variational}. Nevertheless, within the variational updates formulation, the values of $\vb{N}_{n+1}$ and of $\Delta{\gamma}_{n+1}$ are obtained from the minimisation process, and not following some model assumptions.

At the structural domain level, the functional defining the minimisation problem yielding $\vb{u}_{n+1}$ as well as $\Delta{\gamma}_{n+1}\left(\vb{x}\right)$ and $ \vb{N}_{n+1}\left(\vb{x}\right)$, $\forall \vb{x} \in V$, is obtained by combining Eqs. (\ref{eq:energyVar}, \ref{eq:dispMin}, \ref{eq:variationalJ2}, \ref{eq:variationalJ2Constraints}), resulting into
\begin{eqnarray}
	\Delta \Phi_{n+1} = \Delta \Phi_{n+1}\left(\vb{u}_{n+1},\, \Delta\gamma_{n+1},\, \vb{N}_{n+1}\right) \,,
\end{eqnarray}
with the constraints (\ref{eq:variationalJ2Constraints}) having to be applied at every material point $\vb{x} \in V$.
}

\subsection{Finite element approximation}\label{sec:fem}

In a classical finite element approach, the resolution of the material law, \emph{e.g.} the minimisation of Eq. (\ref{eq:energyJ2Inc_body}), is conducted at the local level, \emph{i.e.} at the quadrature points, from an estimation of the displacement field. The latter is corrected considering the resolution of the balance equation at the finite element mesh level. Nested Newton-Raphson loops are thus required at both local and global levels. In this section we summarise the finite element approach in which the resolution of the constitutive law (\ref{eq:energyJ2Inc_body}) is conducted at the finite element level. 

The functional (\ref{eq:energyVar}) is estimated using the finite element approximation. It is necessary to discretise not only the displacement field $\vb{u}$ but also the internal variables field $\vb{q}$. The displacement field degrees of freedom are defined at the finite element nodes whilst the internal variable degrees of freedom are defined at the quadrature points in order to avoid their interpolation, which could be inconsistent, respectively defining the vectors of degrees of freedom $\vb{U}$ and $\vb{Q}$. More details on the finite element discretisation are given in \ref{sec:femApp}.

The energy functional (\ref{eq:energyVar}) is therefore estimated from the fields evaluated at the quadrature points and from the variation of incremental functional $\Delta \mathcal{E}_{n+1}$ evaluated following Section \ref{sec:constitutiveModels}, leading to 
\begin{eqnarray}
	\Delta \Phi_{n+1} &=&  \Delta W^\text{int}_{n+1} - \Delta W^\text{ext}_{n+1}\,, \label{eq:energyVar2_body}
\end{eqnarray}
where
\begin{eqnarray}
	\Delta W^\text{int}_{n+1} &=& \sum_{e=0}^{N_\text{ele}-1}\sum_{\eta=0}^{m_e-1} \omega^{e,\eta} \Delta \mathcal{E}_{n+1}\left( \vb{B}^{e,\eta}\vb{U}^e_{n+1},\,\vb{q}_{n+1}^{e,\eta} \right) \,,\text{ and} \label{eq:WintIncr2_body} \\	
	\Delta W^\text{ext}_{n+1} &=& \Delta \vb{U}_{n+1}^T \mathbf{f}^\text{ext}_{n+1}\,,
\end{eqnarray}
where $N_\text{ele}$ is the number of finite elements, $m_e$ is the number of quadrature points in element $e$, $\vb{B}^{e,\eta}$ is the elementary matrix of the gradient of the shape functions associated with the elementary displacement vector field $\vb{U}^e$ evaluated at the quadrature point $\eta$ of element $e$ --\emph{i.e.} $\bm{\Xi}^{e,\eta}$, $\vb{q}^{e,\eta}$ is the internal variables defined at the quadrature point $\bm{\Xi}^{e,\eta}$, $\omega^{e,\eta}$ denotes the weights of the quadrature points $\bm{\Xi}^{e,\eta}$, and where $\mathbf{f}^\text{ext}_{n+1}$ is the external force vector.

\secondreviewer{The double-minimisation problem detailed in Alg. \ref{alg:DoubleMinimisation} is reformulated as the variational formulation of the finite element method that consists in evaluating the minimum state of the constrained energy functional (\ref{eq:energyVar2_body}), which is convex for the models presented in Section \ref{sec:constitutiveModels}, following
  \begin{eqnarray}\label{eq:doubleMinimisationFEM}
    \begin{cases}
      \vb{Q}_{n+1} &= \argmin{{\vb{Q}_{n+1}^{\prime}}} \Delta\Phi_{n+1}\left(\vb{U}_{n+1},\,\vb{Q}_{n+1}^{\prime}\right)\;\text{ under constraints on } \;\vb{Q}_{n+1}^{\prime}\;; \\
      \Delta\Phi^{\text{eff}}_{n+1}\left(\vb{U}_{n+1}\right)&=\min_{\vb{Q}_{n+1}^{\prime}} \Delta\Phi_{n+1}\left(\vb{U}_{n+1},\,\vb{Q}_{n+1}^{\prime}\right)\;\text{ under constraints on } \;\vb{Q}_{n+1}^{\prime}\;;\\
      \vb{U}_{n+1}  &=  \argmin{{\vb{U}_{n+1}^{\prime}}} \Delta\Phi^{\text{eff}}_{n+1}\left(\vb{U}^{\prime}_{n+1}\right)\;\text{ with } \;\vb{U}^{\prime}_{n+1} \;\text{ kinematically admissible}\,. 
    \end{cases}
  \end{eqnarray}
}

\secondreviewer{The double-minimisation process (\ref{eq:doubleMinimisationFEM}) could be solved with a classical hardware using available optimisation methods. In particular, since the functional arising from the variational formulation is convex, conic programming \cite{Bleyer2020} was shown to be well suited to conduct finite element analyses since both the free energy and dissipation pseudo-potential can be formulated to be conic representable \cite{Bleyer2022}.
  However, conducting the optimisation on a classical computer involves creating algorithms that execute computer instructions sequentially since classical computers process information using bits, which can only be either 0 or 1. On the other hand, quantum annealing leverages principles of quantum mechanics to find solutions to optimisation problems at once by exploiting quantum superposition and entanglement to explore multiple possible solutions simultaneously. This could make the quantum annealing potentially more efficient than classical optimisers in classical computers. This is particularly true since the minimisation problem in terms of the internal variables is simply the combination of independent constrained minimisation problems that do not require all the qubits to be connected and can therefore exploit the parallelism advantage of quantum computers.
Nevertheless not all functions can be encoded into the current quantum annealing hardware since only minimisation of quadratic potentials under the form of either Eq. (\ref{eq:QAdefSpin}) or Eq. (\ref{eq:QAdef}) can be conducted, which motivates the development of a hybrid resolution of the double-minimisation problem.}

\section{Hybrid resolution of the double-minimisation problem}\label{sec:QA-SQP}

This section presents a hybrid strategy, \emph{i.e.} using classical computers and the quantum annealing approach described in Section \ref{sec:qaSummary}, to solve the double-minimisation problem arising from the variational formulation described in Section \ref{sec:mechaBVP}.
For this purpose, we suggest a quantum annealing-assisted sequential quadratic programming procedure to minimise a multivariate convex function with bound constraints.
This procedure can then be used for the resolution of the double-minimisation described in Alg. \ref{alg:DoubleMinimisation}, whenever such a minimisation is required, \emph{i.e.} at lines 4 and 5.

\subsection{Quantum annealing-assisted sequential quadratic programming (QA-SQP)} \label{subsec:QA-SQP}

Sequential Quadratic Programming (SQP) is a numerical optimisation method used to solve non-linear optimisation problems.
It is an iterative algorithm that aims at finding the solution of a multivariate optimisation problem by approximating it as a series of quadratic problems.

Let us consider a multivariate minimisation problem of the $N$-variable objective function $f$:
\begin{eqnarray}\label{eq:unconstrainedMinimistionDefinition}
	\begin{cases}
		&\min_{\vb{w}} f\left(\vb{w}\right)\,,\\
		&\text{subject to } \vb{w}^{\text{min}}\leq \vb{w}\leq \vb{w}^{\text{max}} \,.
	\end{cases}
\end{eqnarray}
\secondreviewer{For the cases involving the following equality and/or inequality constraints
\begin{eqnarray}\label{eq:optDefConstrained}
\begin{cases}
		h_j\left(\vb{w}\right) &= 0 \text{ with } j =0,\ldots, N_h-1 \,, \text{ and }\\
		l_j\left(\vb{w}\right) &\leq 0 \text{ with } j =0,\ldots, N_l-1 \,, \\
	\end{cases} 
\end{eqnarray}
where $N_h$ and $N_l$ are the number of equality and inequality constraints, respectively, the corresponding constrained minimisation problem needs to be transformed into the quasi-unconstrained form (\ref{eq:unconstrainedMinimistionDefinition}). For this purpose, we define the augmented objective function using penalties as
\begin{eqnarray}\label{eq:augmentedFunc}
f_{\text{aug}}\left(\vb{v}\right)=f_{\text{aug}}\left(\vb{w},\,\bm{\lambda}\right)= f\left(\vb{w}\right) + \sum_{j=0}^{N_h-1} c_j^h \left[h_j\left(\vb{w}\right)\right]^2 +  \sum_{j=0}^{N_l-1} c_j^l \left[l_j\left(\vb{w}\right) + \lambda_j\right]^2 \,,
\end{eqnarray}
where $c_j^h$ and $c_j^l$, $\forall j$ are the penalty factors associated with the constraints $h_j$ and $l_j$, respectively, $\bm{\lambda}=\left[\lambda_j\,,\forall j=0,\ldots,N_l-1\right]$ is the vector of the so-called auxiliary variables, and where $\vb{v}=\left[\vb{w}^T\;\bm{\lambda}^T\right]^T$ gathers all the unknowns.
The quasi-unconstrained form (\ref{eq:unconstrainedMinimistionDefinition}) is then rewritten as
\begin{eqnarray}\label{eq:constrainedMinimistionDefinitiontmp}
	\min_{\vb{w}, \bm{\lambda}} f_{\text{aug}}\left(\vb{w},\,\bm{\lambda}\right) \;
	\text{ subject to }  \vb{w}^{\text{min}}\leq \vb{w}\leq \vb{w}^{\text{max}} \text{ and } \lambda_{j} \geq 0\,\, \forall  j =0,\ldots, N_l-1 \,,
\end{eqnarray}
or in the condensed form
\begin{eqnarray}\label{eq:constrainedMinimistionDefinition}
	\min_{\vb{v}} f_{\text{aug}}\left(\vb{v}\right) \;
	\text{ subject to }  \vb{v}^{\text{min}}\leq \vb{v}\leq \vb{v}^{\text{max}} \,,
\end{eqnarray}
in which the lower and upper bounds associated to the components of $\vb{v}$ corresponding to $\lambda_j$ are respectively $0$ and $\infty$.}

For a commonly encountered situation in computational mechanics in which the gradient $\vb{g}$ and the non-negative definite hessian $\vb{K}$ of $f_{\text{aug}}$ exist everywhere, one has the following approximation
\begin{eqnarray}\label{eq:quadApprox}
	f_{\text{aug}}\left(\vb{v} + \vb{z} \right)  \approx f_{\text{aug}}\left(\vb{v}\right) + \text{QF}\left(\vb{z};\,\vb{g},\, \vb{K} \right)\,,
\end{eqnarray}
where the components of $\vb{g}$ and of $\vb{K}$ read
\begin{eqnarray}
	g_i = \eval{\frac{\partial f_{\text{aug}}}{\partial v_i}}_{\vb{v}} \text{ and } {K}_{ij} = \eval{\frac{\partial^2 f_{\text{aug}}}{\partial v_i \partial v_j}}_{\vb{v}} \label{eq:gradienthess}\,,
\end{eqnarray} 
and where $\text{QF}\left(\vb{z};\,\vb{g},\, \vb{K} \right)$ denotes a general quadratic form in the vector $\vb{z}$ defining the problem unknowns.
This quadratic form is defined by the vector $\vb{g}$ and the matrix $\vb{K}$ following
\begin{eqnarray}\label{eq:quadForm}
	\text{QF}\left(\vb{z};\,\vb{g},\,\vb{K}\right)=\vb{z}^T\vb{g} + \frac{1}{2}\vb{z}^T\vb{K}\vb{z}\,.
\end{eqnarray}
The key steps in the SQP algorithm to solve the minimisation problem stated by Eq. (\ref{eq:constrainedMinimistionDefinition}) are:
\begin{itemize}
	\item[(i)] Start with an initial admissible solution;
	\item[(ii)] Construct a quadratic model (\ref{eq:quadForm}) of the objective function $f$ and the constraints around the current solution;
	\item[(iii)] Solve the quadratic minimisation problem $\vb{z}^* = \argmin{\vb{z}} \text{QF}\left(\vb{z};\,\vb{g},\, \vb{K}\right)$ with respect to the constraints of Eq. (\ref{eq:constrainedMinimistionDefinition});
	\item[(iv)] Check a convergence criterion: if the convergence is reached terminate the algorithm; otherwise, go back to step (ii).
\end{itemize}
In this work, we suggest to use the quantum annealing for step (iii), leading to the so-called Quantum Annealing-assisted Sequential Quadratic Programming (QA-SQP). For this purpose, the transformation from the continuous quadratic form (\ref{eq:quadForm}) into the binary quadratic form (\ref{eq:QAdef}) is first introduced. Then a hybrid nested scheme to solve the minimisation problem (\ref{eq:constrainedMinimistionDefinition}) is provided.

\subsubsection{Conversion from QF to QUBO}

A $L$-bit string, \emph{i.e.} ${b_{L-1}\ldots b_{0}}$ with $b_i \in \{0,1\}$, can represent an arbitrary whole number ranging from 0 to $2^{L}-1$ through the binary-decimal conversion
\begin{eqnarray}
	{b_{L-1}\ldots b_{0}} \equiv \sum_{j=0}^{L-1} b_j 2^j\,.
\end{eqnarray}
As a result, one can discretise the range $\mqty[z^{\text{min}},\,z^{\text{max}}]$ into $2^{L}$ discrete values following
\begin{eqnarray}\label{eq:binarised}
	z\left(b_0,\,\ldots,\,b_{L-1}\right) = \bar{z} + \epsilon \left( \sum_{j=0}^{L-1} b_j 2^j  - 2^{L-1}+1\right) \in \mqty[z^{\text{min}},\,z^{\text{max}}]\,,
\end{eqnarray}
where $\epsilon$ is the so-called discretisation error, and where 
\begin{eqnarray}
	\begin{cases}
		z^{\text{min}} &= \bar{z}- \left( 2^{L-1}-1\right) \epsilon\,,  \text{ and}\\
		z^{\text{max}}&= \bar{z} +2^{L-1} \epsilon\,.
	\end{cases}
\end{eqnarray}
Therefore, Eq. (\ref{eq:binarised}) can be rewritten as
\begin{eqnarray}\label{eq:binarisationVec}
	z\left(\vb{b}\right)=  z^{\text{min}} + \epsilon \bm{\beta}^T\vb{b}\,,
\end{eqnarray}
where $\bm{\beta} = \mqty[2^0&2^1\,.\ldots&2^{L-1}]^T$ and $\vb{b}=\mqty[b_0&\ldots&b_{L-1}]^T$. For a given error $\epsilon$, a larger range $\mqty[z^{\text{min}},\,z^{\text{max}}]$ is obtained using more qubits whilst when the number of qubits is fixed, the range $\mqty[z^{\text{min}},\,z^{\text{max}}]$ becomes smaller with decreasing $\epsilon$.

Let us consider a vector $\vb{z}$ of $N$ components, Eq. (\ref{eq:binarisationVec}) is thus applied to each component of $\vb{z}$, leading to
\begin{eqnarray}\label{eq:binaryApprox}
	\vb{z} = \vb{z}^{\text{min}} + \left[\epsilon_i\bm{\beta}^T\vb{b}_i \text{ for } i=0,\,\ldots,\, N-1\right] = \vb{a} + \vb{D}\vb{b} \in \mqty[\vb{z}^{\text{min}},\,\vb{z}^{\text{max}}]\,,
\end{eqnarray}
where $\bullet_i$ denotes a quantity associated with the component $z_i$, $\vb{z}^{\text{min}}$ and $\vb{z}^{\text{max}}$ denote respectively the lower-bound and upper-bound of $\vb{z}$ and where
\begin{eqnarray}
	\vb{a}&=&\vb{z}^{\text{min}} = \mqty[z_0^\text{min}&\ldots&z_{N-1}^\text{min}]^T \label{eq:binaryApproxa}\,,\\
	\vb{D} &=&\text{diag}\left(\epsilon_0\bm{\beta}^T,\,\ldots,\,\epsilon_{N-1}\bm{\beta}^T \right) \,, \text{ and }\label{eq:binaryApproxD}\\
	\vb{b} &=& \mqty[\vb{b}_0^T&\ldots\,\vb{b}_{N-1}^T]^T\,,\label{eq:binaryApproxb}
\end{eqnarray}
with $\text{diag}\left(\bullet\right)$ denoting the $N\times\left(N\times L\right)$ block diagonal matrix. 

We now reconsider the multivariate optimisation problem in terms of the objective function (\ref{eq:quadApprox}). Using Eq. (\ref{eq:binaryApprox}), Eq. (\ref{eq:quadForm}) becomes
\begin{eqnarray}\label{eq:QFApp2}
	\text{QF}\left(\vb{z};\,\vb{g},\,\vb{K} \right) &=&  \frac{1}{2} \vb{b}^T \vb{D}^T \vb{K} \vb{D} \vb{b} + \vb{b}^T\vb{D}^T \left(\vb{K} \vb{a} +\vb{g}\right) + \frac{1}{2} \vb{a}^T \vb{K}\vb{a} + \vb{a}^T \vb{g}\,.
\end{eqnarray}
Since $b_i\in\{0,1\}$, one has $b_i^2 = b_i\,,\;\forall i$. As a result, Eq. (\ref{eq:QFApp2}) can be rewritten in terms of a QUBO function (\ref{eq:QAdef}) as
\begin{eqnarray} \label{eq:QFQUBOEquiv}
	\text{QF}\left(\vb{z};\,\vb{g},\,\vb{K} \right) = \text{QUBO}\left(\vb{b};\, \vb{A}\right) + \frac{1}{2} \vb{a}^T \vb{K}\vb{a} + \vb{a}^T \vb{g}\,,
\end{eqnarray}
where 
\begin{eqnarray}\label{eq:quboMatrix}
	\vb{A} = \frac{1}{2}\vb{D}^T \vb{K} \vb{D}  + \text{diag}\left(\vb{D}^T \left(\vb{K} \vb{a} +\vb{g}\right) \right)\,,
\end{eqnarray}
with $\text{diag}\left(\bullet\right)$ denoting the diagonal matrix whose diagonal is given by a vector $\bullet$.

Contrarily to the quadratic form (\ref{eq:quadForm}), which is expressed in terms of continuous variables, the binary quadratic form (\ref{eq:QFQUBOEquiv}) can be minimised by quantum annealing and the continuous variables are recovered using Eq. (\ref{eq:binaryApprox}).
The difficulty is now to define the parameters $\vb{a}$ and $\vb{D}$ required for the binary encoding (\ref{eq:binaryApprox}). Toward this end, an iterative procedure is introduced in the hybrid nested scheme.

\subsubsection{Hybrid nested scheme}

\begin{figure}[!htb]
	\centering
	\includegraphics[scale=0.5,valign=c]{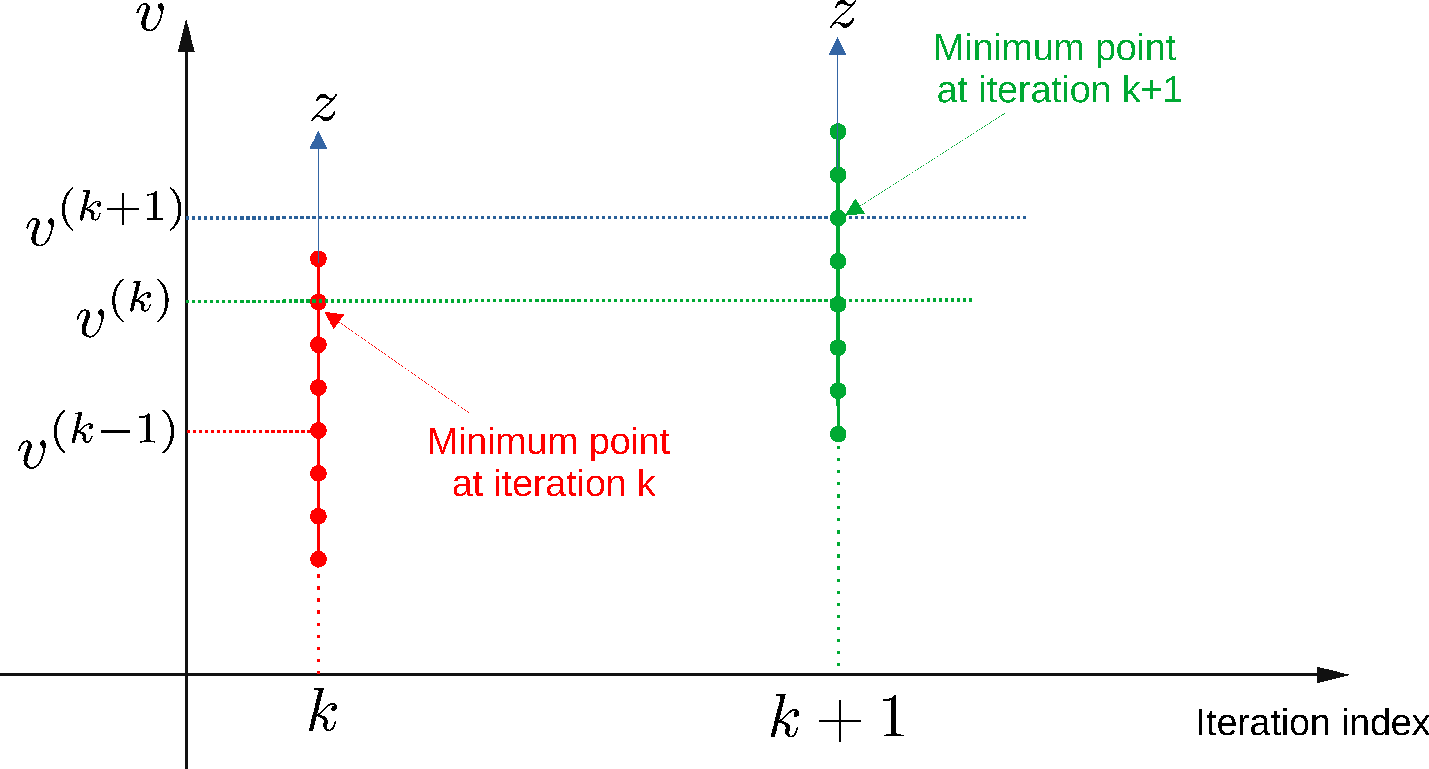}
	\caption{Illustration of the minimisation of a single variable $v$ function considering $L=3$ qubits. At each iteration, the minimum between 8 discrete values is found and is used as the starting point for the next iteration.}
	\label{fig:oneDMiniStep}
\end{figure}

The basic idea of the iterative procedure is illustrated in Fig. \ref{fig:oneDMiniStep} when performing the minimisation of a function of a single variable $v$ with $L=$3 qubits. At an iteration $k$, the solution $v^{(k-1)}$ and the discretisation error $\epsilon^{(k-1)}$ are known.
Using $L$ qubits, the discretisation error $\epsilon^{(k)}=\epsilon^{(k-1)}$, and $\bar{z}^{(k)}=0$, $2^L$ discrete values are defined around $v^{(k-1)}$.
As a result a QUBO problem is defined as shown by Eq. (\ref{eq:QFQUBOEquiv}) and whose minimisation allows obtaining $v^{(k)}$ for the next iteration.
To this end, two situations need to be considered:
\begin{itemize}
	\item At iteration $k$, the upper-bound and lower-bound of $v$ have to be respected, \emph{i.e.}
	\begin{eqnarray}
		v^{\text{min}}\leq v^{(k-1)} + z \leq v^{\text{max}}\,.
	\end{eqnarray}  
	For this purpose, the discretisation around the solution of the previous iteration needs to be adjusted as follows
	\begin{eqnarray}
		z^{\text{min}} &=& \max\left(v^{\text{min}}-v^{(k-1)},\, -(2^{L-1}-1)\epsilon^{(k-1)}\right)\,,\\
		z^{\text{max}} &=& \min\left(v^{\text{max}}-v^{(k-1)},\, 2^{L-1}\epsilon^{(k-1)}\right)\,,\\
		\epsilon^{(k)} &=& \cfrac{z^{\text{max}} - z^{\text{min}}}{2^L-1} \label{eq:errorDis}\,,\text{ and }\\
		\bar{z}^{(k)} &=&  \cfrac{z^{\text{max}} + z^{\text{min}} -\epsilon^{(k)} }{2}\,. \label{eq:newzbar}
	\end{eqnarray}
	Finally, the new discretisation is built around $\bar{z}^{(k)}$ following Eq. (\ref{eq:newzbar}) and using the discretisation error $\epsilon^{(k)}$ following Eq. (\ref{eq:errorDis}).
	\item The objective function value does not decrease. In this situation, the discretisation error is decreased by a factor $\xi < 1$, \emph{i.e.} $\epsilon^{(k)} = \epsilon^{(k)} \xi$. In this work, we use $\xi=0.5$. 
\end{itemize}      

\begin{algorithm}
	\caption{QA-SQP for solving the minimisation problem stated by Eqs. (\ref{eq:constrainedMinimistionDefinition}).}
	\label{alg:QA}
	\begin{algorithmic}[1]
		\Require{$\vb{v}_0$ (initial solution), $L$ (number of qubits), $\vb{z}^{\text{min}}$ (lower bound), $\vb{z}^{\text{max}}$ (upper bound), $\bm{\epsilon}_0$ (initial discretisation error), $\xi$ (shrinking factor)}, $N_\text{steps}$ (maximum number of successful iterations), $N_\text{failed}$ (maximum number of failed iterations)
		\Ensure{$\vb{v}$ (optimised solution)}
		\State initialise $\vb{v}\gets \vb{v}_0$, $k \gets 0$, $l\gets 0$;
		\State estimate $f_{\text{aug}}$ following Eq. (\ref{eq:augmentedFunc}) and  $\vb{g}$, $\vb{K}$ following Eq. (\ref{eq:gradienthess}) at $\vb{v}$;
		\State initialise $\bm{\epsilon}\gets\bm{\epsilon}_0$;
		\Repeat
		\State update $\bm{\epsilon}$ and $\bar{\vb{z}}$ from $L$, $\vb{v}$, $\vb{z}^{\text{min}}$, and $\vb{z}^{\text{max}}$ following Eq. (\ref{eq:updateBounds});
		\State estimate $\vb{a}$ and $\vb{D}$ from $L$, $\bar{\vb{z}}$ and $\bm{\epsilon}$ following Eqs. (\ref{eq:binaryApproxa}, \ref{eq:binaryApproxD});
		\State estimate the QUBO matrix $\vb{A}$ following Eq. (\ref{eq:quboMatrix});
		\State call quantum annealer (\ref{eq:QAdef}), leading to the solution $\vb{b}$;
		\State estimate $\vb{z} = \vb{a} +\vb{D} {\vb{b}}$;
		\State estimate $f_{\text{aug}_\text{new}} = f_{\text{aug}}\left(\vb{v}+\vb{z}\right)$ following Eq. (\ref{eq:augmentedFunc});
		\If {$f_{\text{aug}_\text{new}} < f_{\text{aug}}$} 
		\State $k \gets k+1$;
		\State estimate $\vb{g}_\text{new}$, $\vb{K}_\text{new}$ at $\vb{v}+\vb{z}$ following Eq. (\ref{eq:gradienthess});
		\State $\vb{v}\gets\vb{v}+\vb{z} $, $f_{\text{aug}}\gets f_{\text{aug}_\text{new}}$, $\vb{g}\gets\vb{g}_\text{new}$, and $\vb{K}\gets\vb{K}_\text{new}$;
		\Else
		\State $l\gets l+1$;
		\State $\bm{\epsilon} \gets  \max(\bm{\epsilon}^{\text{min}}, \xi \bm{\epsilon})$;
		\EndIf
		\If {$k \geq N_\text{steps}$ or $l \geq N_\text{failed}$}
		\State break;
		\EndIf
		\Until{convergence}
	\end{algorithmic}
\end{algorithm}
The minimisation problem (\ref{eq:quadApprox}) of the multi-variable function is directly obtained by generalisation of the single variable case above and is detailed in Alg. \ref{alg:QA}.
The solution is initialised with a feasible initial guess $\vb{v}_0$, \emph{i.e.} $ \vb{v}^{\text{min}}\leq \vb{v}_0\leq \vb{v}^{\text{max}}$. This allows initialising the values of $\vb{g}$ and $\vb{K}$ in line (2). In line (5), the pair ($\bm{\epsilon}$, $\bar{\vb{z}}$) defines a range for the solution of the quadratic form (\ref{eq:quadForm}). The initial discretisation error $\bm{\epsilon}_0$ can be chosen through an initial solution interval $\Delta \vb{v}_0$ such that 
\begin{eqnarray}
	\bm{\epsilon}_0 &= \cfrac{\Delta \vb{v}_0}{2^L-1} \,.
\end{eqnarray}
At each iteration with the current solution $\vb{v}$, the value of $\vb{z}$ will be found such that
\begin{eqnarray}
	\vb{v}^{\text{min}}- \vb{v}\leq \vb{z} \leq \vb{v}^{\text{max}}-\vb{v}\,.
\end{eqnarray}
Besides, for the current $\bm{\epsilon}$ and $\bar{\vb{z}}=\vb{0}$, we have the range of the solution
\begin{eqnarray}
	\vb{v}- \left( 2^{L-1}-1\right) \bm{\epsilon}\leq \vb{v}+\vb{z} \leq \vb{v} +2^{L-1} \bm{\epsilon}\,,
\end{eqnarray}
leading to new bounds on $\vb{z}$,
\begin{eqnarray} 
	\begin{cases}
		\vb{z}^{\text{min}} &= \max(\vb{v}^{\text{min}}- \vb{v},\, - \left( 2^{L-1}-1\right) \bm{\epsilon})\,,\text{ and}\\
		\vb{z}^{\text{max}} &= \min(\vb{v}^{\text{max}}- \vb{v},\, 2^{L-1} \bm{\epsilon}) \,.
	\end{cases}
\end{eqnarray}
As a result, the updated values of the pair ($\bm{\epsilon}$, $\bar{\vb{z}}$) read
\begin{eqnarray}\label{eq:updateBounds}
	\begin{cases}
		\bm{\epsilon} &= \cfrac{\vb{z}^{\text{max}} -\vb{z}^{\text{min}}}{2^L-1} \,,\text{ and}\\
		\bar{\vb{z}}&=\cfrac{\vb{z}^{\text{min}}+ \vb{z}^{\text{max}}-\epsilon}{2}\,.
	\end{cases}
\end{eqnarray}

\subsection{Application of the QA-SQP to the finite element double-minimisation}\label{subsec:doubleMinFE}

In this section, the QA-SQP algorithm proposed in the previous section is particularised to solve the double-minimisation problem detailed in Alg. \ref{alg:DoubleMinimisation} arising from the variational formulation of the finite element method.

\begin{algorithm}[htb!]
	\caption{Finite element simulation with Double-Minimisation Algorithm.}
	\label{alg:DoubleMinimisationFEM}
	\begin{algorithmic}[1]
		\Require $\vb{U}_{n}$, $\vb{Q}_n$;
		\Ensure $\vb{U}_{n+1}$, $\vb{Q}_{n+1}$;
		\State initialise ${\vb{U}}_{n+1}\gets\vb{U}_n$ and $\vb{Q}_{n+1}\gets\vb{Q}_n$;
		\State apply boundary conditions;
		\Repeat
		\State $\vb{U}_{n+1}\gets$ minimise the functional (\ref{eq:energyVar2_body}) with the last value of $\vb{Q}_{n+1}$ using QA-SQP, Alg. \ref{alg:QA};
		\State $\vb{Q}_{n+1}\gets$ minimise the functional (\ref{eq:energyVar2_body}) with the last value of $\vb{U}_{n+1}$ using QA-SQP, Alg. \ref{alg:QA};
		\Until{convergence}
	\end{algorithmic}
\end{algorithm}

Following Section \ref{sec:fem} and the set of Eqs. (\ref{eq:doubleMinimisationFEM}), the finite element simulation version of Alg. \ref{alg:DoubleMinimisation} follows as Alg. \ref{alg:DoubleMinimisationFEM}.
As a reminder, the vectors of degrees of freedom $\vb{U}$ and $\vb{Q}$ arise from the discretisation of the displacement field $\vb{u}$ and of the internal variables field $\vb{q}$, respectively.

\subsubsection{Constitutive model quasi-constrained minimisation}\label{subsec:doubleMinJ2}

\secondreviewer{In the case of $J_2$-plasticity, the optimisation problem stated in line (5) in Alg. \ref{alg:DoubleMinimisationFEM} corresponds to the constrained minimisation problem stated by Eqs. (\ref{eq:variationalJ2}, \ref{eq:variationalJ2Constraints}). To apply the framework developed in Section \ref{subsec:QA-SQP}, this constrained minimisation problem needs to be transformed into a quasi-constrained minimisation of the form (\ref{eq:augmentedFunc}-\ref{eq:constrainedMinimistionDefinition}). For this purpose, we minimise the augmented incremental functional defined from Eq. (\ref{eq:energyJ2Inc_body}) as 
\begin{eqnarray}\label{eq:constrainedJ2}
\Delta{\gamma}_{n+1}, \bm{\alpha}=	&&\argmin{\Delta{\gamma}_{n+1}, \bm{\alpha}}  \left[\Delta\mathcal{E}_{n+1}\left(\vb{u}_{n+1},\,\Delta{\gamma}_{n+1},\,\vb{N}_{n+1}\right) + c_0^h \left(\bm{\alpha}^T \vb{M}\bm{\alpha}-\cfrac{3}{2} \right)^2 \right]\,, \\
	&&\text{subject to } \Delta{\gamma}_{n+1} \geq 0\,, \text{ and } -\sqrt{\cfrac{3}{2}}\leq\alpha_j \leq \sqrt{\cfrac{3}{2}} \;\; \forall \alpha_j \in \bm{\alpha} \nonumber\,,
\end{eqnarray}
where $\vb{N}_{n+1}$ is parameterised as a purely deviatoric tensor such that $\vb{N}_{n+1}:\vb{N}_{n+1} = \bm{\alpha}^T\vb{M}\bm{\alpha}$ with\footnote{In a two-dimensional setting, one has $\alpha_3=\alpha_4=0$, leading to
\begin{eqnarray}
	\vb{N}_{n+1} = \mqty[\alpha_0&\alpha_2/\sqrt{2}&0\\\alpha_2/\sqrt{2}&\alpha_1&0\\0&0&-\alpha_0-\alpha_1], \bm{\alpha}=\mqty[\alpha_0\\\alpha_1\\\alpha_2], \text{ and } \vb{M}=\mqty[2&1&0\\1&2&0\\0&0&1]\nonumber\,.
\end{eqnarray}}
\begin{eqnarray}
	\vb{N}_{n+1} = \mqty[\alpha_0&\alpha_2/\sqrt{2}&\alpha_3/\sqrt{2}\\\alpha_2/\sqrt{2}&\alpha_1&\alpha_4/\sqrt{2}\\\alpha_3/\sqrt{2}&\alpha_4/\sqrt{2}&-\alpha_0-\alpha_1], \bm{\alpha}=\mqty[\alpha_0\\\alpha_1\\\alpha_2\\\alpha_3\\\alpha_4], \text{ and } \vb{M}=\mqty[2&1&0&0&0\\1&2&0&0&0\\0&0&1&0&0\\0&0&0&1&0\\0&0&0&0&1]\,.
\end{eqnarray}
The penalty parameter $c_0^h$ is choose to be equal to the shear modulus $\mu$  in Eq. (\ref{eq:constrainedJ2}).

In practice, each minimisation problem of Alg. \ref{alg:DoubleMinimisationFEM} is performed using the QA-SQP presented in Alg. \ref{alg:QA}. Actually, Eq. (\ref{eq:constrainedJ2}) is a fourth-order polynomial in terms of $\bm{\alpha}$. When using the binarisation (\ref{eq:binaryApprox}), the higher third- and fourth-order binary terms arise, leading to a higher-order unconstrained binary optimisation (HUBO) problem in terms of $\bm{\alpha}$. Such a HUBO can be transformed into a QUBO by replacing the product of two binary variables by a new one and by including an appropriate penalty term to enforce this equality \cite{rosenberg1975reduction, boros2002pseudo,jiang2018quantum,mandal2020compressed}. In that case the constraints are naturally quadratic and the quadratic approximation (\ref{eq:quadApprox}) is no longer required to handle the constraints part of the objective function (\ref{eq:augmentedFunc}). However, in this work, since we have to consider this quadratic approximation for the other terms of the objective function anyway, we apply the developed SQP strategy for all the terms of the augmented incremental functional defined in Eq. (\ref{eq:constrainedJ2}).}

\subsubsection{Finite element assembly}\label{subsec:doubleMinAssembly}

In order to make the finite element optimisation formulation ready for QA-SQP, the quadratic approximation of the function (\ref{eq:energyVar2_body}) needs to be constructed.

For a given $\Delta\vb{U}_{n+1}$ and a given $\Delta\vb{Q}_{n+1}$, the quadratic approximation of Eq. (\ref{eq:energyVar2_body}) with respect to the perturbation of $\Delta\vb{Q}_{n+1}$, \emph{i.e.} $\vb{z}_Q$, reads
\begin{eqnarray}
	\Delta \Phi_{n+1}\left(\Delta\vb{U}_{n+1}, \Delta\vb{Q}_{n+1}+\vb{z}_Q \right) \approx \Delta \Phi_{n+1}\left(\Delta\vb{U}_{n+1}, \Delta\vb{Q}_{n+1}\right) +  \vb{z}_Q^T \vb{r}_{n+1} + \frac{1}{2} \vb{z}_Q^T \vb{T}_{n+1} \vb{z}_Q\,, 
\end{eqnarray}
where
\begin{eqnarray}
	\vb{r}_{n+1} &=& \cfrac{\partial \Delta \Phi_{n+1}}{\partial \Delta\vb{Q}_{n+1}} = -\assembleOperator_{V^e\subset V}\sum_{\eta=0}^{m_e-1} \omega^{e,\eta} \vb{Y}_{n+1}\,, \text{ and}\\
	\vb{T}_{n+1} &=& \cfrac{\partial \Delta \vb{r}_{n+1}}{\partial \Delta\vb{Q}_{n+1}} = -\assembleOperator_{V^e\subset V}\sum_{\eta=0}^{m_e-1} \omega^{e,\eta} \vb{Z}_{n+1}\,,
\end{eqnarray}
with $\assembleOperator$ representing the assembly process of the elements $V^e$ of the volume $V$, and with $ \vb{Y}_{n+1}$ and $\vb{Z}_{n+1}$ being available from $\Delta \mathcal{E}_{n+1}$ as 
\begin{eqnarray}
	\vb{Y}_{n+1} = -\frac{\partial\Delta \mathcal{E}_{n+1}}{\partial\Delta\vb{q}_{n+1}} \text{ and } \vb{Z}_{n+1} = \frac{\partial  \vb{Y}_{n+1}}{\partial\Delta\vb{q}_{n+1}}\,.
\end{eqnarray}

For a given $\Delta\vb{U}_{n+1}$ and a given $\Delta\vb{Q}_{n+1}$, the quadratic approximation of Eq. (\ref{eq:energyVar2_body}) with respect to the perturbation of $\Delta\vb{U}_{n+1}$, \emph{i.e.} $\vb{z}_U$, reads
\begin{eqnarray}
	\Delta \Phi_{n+1}\left(\Delta\vb{U}_{n+1}+\vb{z}_U, \Delta\vb{Q}_{n+1}\right) \approx \Delta \Phi_{n+1}\left(\Delta\vb{U}_{n+1}, \Delta\vb{Q}_{n+1}\right) +  \vb{z}_U^T \vb{g}_{n+1} + \frac{1}{2} \vb{z}_U^T \vb{K}_{n+1} \vb{z}_U\,, 
\end{eqnarray}
where
\begin{eqnarray}
	\vb{g}_{n+1} &=&\cfrac{\partial \Delta \Phi_{n+1}}{\partial \Delta\vb{U}_{n+1}} = \assembleOperator_{V^e\subset V} \sum_{\eta=0}^{m_e-1} \omega^{e,\eta} \left(\vb{B}^{e,\eta}\right)^T \mqty[\bm{\sigma}_{n+1}] - \mathbf{f}_{n+1}^{\text{ext}} \,,  \text{ and}\\
	\vb{K}_{n+1} &=&\frac{\partial \vb{g}_{n+1}}{\partial \Delta\vb{U}_{n+1}}= \assembleOperator_{V^e\subset V} \sum_{\eta=0}^{m_e-1} \omega^{e,\eta} \left(\vb{B}^{e,\eta}\right)^T \mqty[\vb{C}_{n+1}]  \vb{B}^{e,\eta} \label{eq:stiffness}\,,
\end{eqnarray}
with $\bm{\sigma}_{n+1}$ being the current stress tensor and $\vb{C}_{n+1} $ being the current elastic tangent operator\footnote{Since the internal state is fixed, the material behaviour is considered as an elastic medium during this step of the double-minimisation problem.}
\begin{eqnarray}
	\bm{\sigma}_{n+1} = \frac{\partial \Delta \mathcal{E}^{\text{eff}}_{n+1} }{\partial \Delta\bm{\varepsilon}_{n+1}}\,, \text{ and } \vb{C}_{n+1} = \frac{\partial^2 \Delta \mathcal{E}^{\text{eff}}_{n+1}}{\partial \bm{\varepsilon}^2}=\vb{C}^{\text{el}}_{n+1}\,.
\end{eqnarray}

\section{Numerical examples}\label{sec:examples}

The minimisation algorithm sketched in Alg. \ref{alg:QA} is carried out on the D-wave systems. The binary optimisation problem (\ref{eq:QAdef}) is submitted using the D-wave Ocean Python library \cite{dwaveDoc}. Whenever the matrix $\vb{A}$ in Eq. (\ref{eq:QAdef}) is available, Alg. \ref{alg:submitDwave} shows the Python code to submit the binary optimisation problem to the D-wave quantum annealer. \emph{DWaveSampler} is used to submit the problem to an available quantum processing unit. Since the submitted problem does not automatically fit the hardware graph, it is necessary to find an embedding for this purpose: we consider the general class \emph{EmbeddingComposite} for embedding. Unless stated otherwise, the function \emph{sample\_qubo} is used with 100 reads to evaluate the minimum of the objective function and 20 $\mu$s for the annealing time.
The output is chosen as the variables making the objective function the smallest.
\begin{algorithm}[htb!]
	\centering
	\begin{varwidth}{\linewidth}
		\begin{verbatim}
			from dwave.system import DWaveSampler, EmbeddingComposite
			sampler = EmbeddingComposite(DWaveSampler())
			sampleset = sampler.sample_qubo(A, num_reads=100, annealing_time=20)
			b = sampleset.first.sample
		\end{verbatim}
	\end{varwidth}
	\caption{Binary optimisation problem submitted to the D-wave system.}
	\label{alg:submitDwave}
\end{algorithm}

In what follows, we present the numerical results obtained for elasto-plastic one and two-dimensional problems to demonstrate the applicability of the proposed framework to solve non-linear history-dependent mechanical problems.

\subsection{Elasto-plastic bar under tensile loading}\label{subsec:1DTest}

In this section, we apply the proposed hybrid framework on a uniaxial-strain tensile test, in order to study the effect of the number of qubits used in the binary discretisation.

The displacement field over a bar of length $l$ = 1mm reads
\begin{eqnarray}
	\vb{u}=\mqty[u_x&0&0]^T \text{ over } 0 \leq x \leq l\,.
\end{eqnarray}
The bar is clamped at $x=0$. The governing equations (\ref{eq:balance1}-\ref{eq:balance3}) simplify for this case as
\begin{eqnarray}
	\begin{cases}
		\frac{\dd{\sigma_{xx}}}{\dd x} + b_0&=0 \,, \label{eq:1Dbalance} \\
		u_x\left(x=0\right) &= 0 \,,\text{ and } \\
		\sigma_{xx}\left(x=l\right) &= 0\,, \label{eq:1DforceBC}
	\end{cases}
\end{eqnarray}
where $b_0 >0$ is constant over the bar. The strain tensor $\bm{\varepsilon}$ and the stress tensor $\bm{\sigma}$ over the bar read
\begin{eqnarray}
	\bm{\varepsilon} = \mqty[\varepsilon_{xx}&0&0\\0&0&0\\0&0&0]\,\text{ and } \bm{\sigma} = \mqty[\sigma_{xx}&0&0\\0&\sigma_{yy}&0\\0&0&\sigma_{zz}]
\end{eqnarray}
where ${\varepsilon}_{xx}=\cfrac{\dd u_x }{\dd x} $ and $\sigma_{yy}=\sigma_{zz}$. For a given uniform $b_0$, the analytical solution can be found, see \ref{app:UniaxialTestAnalytical} for details. The material in the bar obeys $J_2$-elasto-plasticity with linear isotropic hardening, and with the material parameters reported in Tab. \ref{tab:materialParameters1}. In this case, the plastic deformation does not occur inside the bar if 
\begin{eqnarray}
	b_0 \leq \left(K+ \cfrac{4}{3}\mu\right)\cfrac{\sigma_y^0}{2\mu l} = 122.5 \,\, \text{MPa}\cdot \text{mm}^{-1} \,.
\end{eqnarray}
\begin{table}[htb]
	\centering
	\caption{Material properties considered in the uniaxial-strain tensile test.}
	\label{tab:materialParameters1}
	\begin{tabular}{|l|c|c|c|}
		\hline
		Parameter&Notation/Expression&Value/Expression&Unit\\
		\hline
		Young's modulus&$E$&2&GPa\\
		\hline
		Poisson's ratio&$\nu$&0.3&-\\
		\hline
		Initial yield stress &$\sigma_y^0$&70&MPa\\
		\hline
		Linear isotropic hardening&$R$&$20\gamma$&MPa\\
		\hline
	\end{tabular}
\end{table}

\begin{figure}[!htb]
	\centering
	\begin{tabular}{cc}
		\includegraphics[scale=0.5,valign=c]{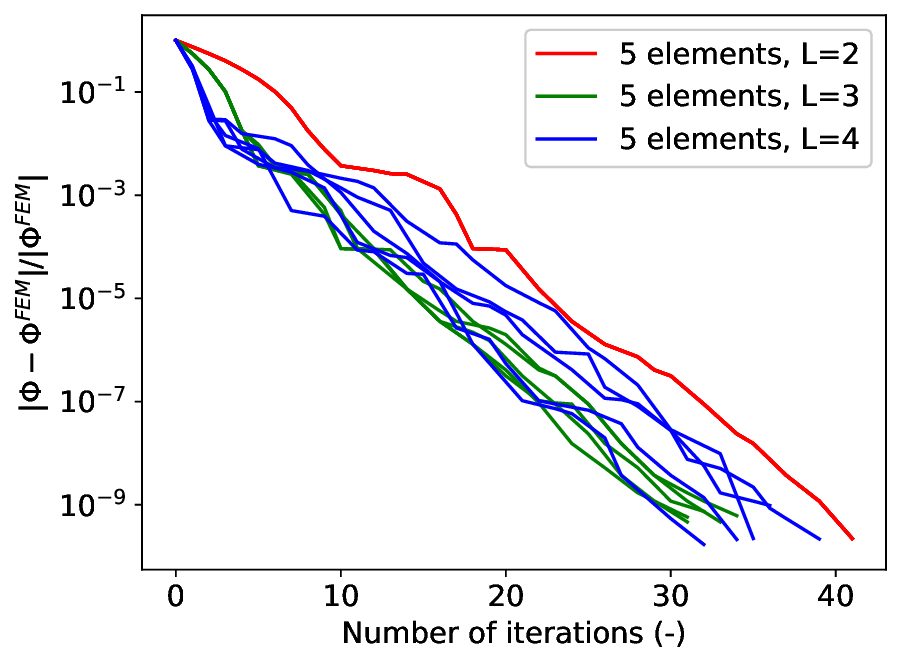}&\includegraphics[scale=0.5,valign=c]{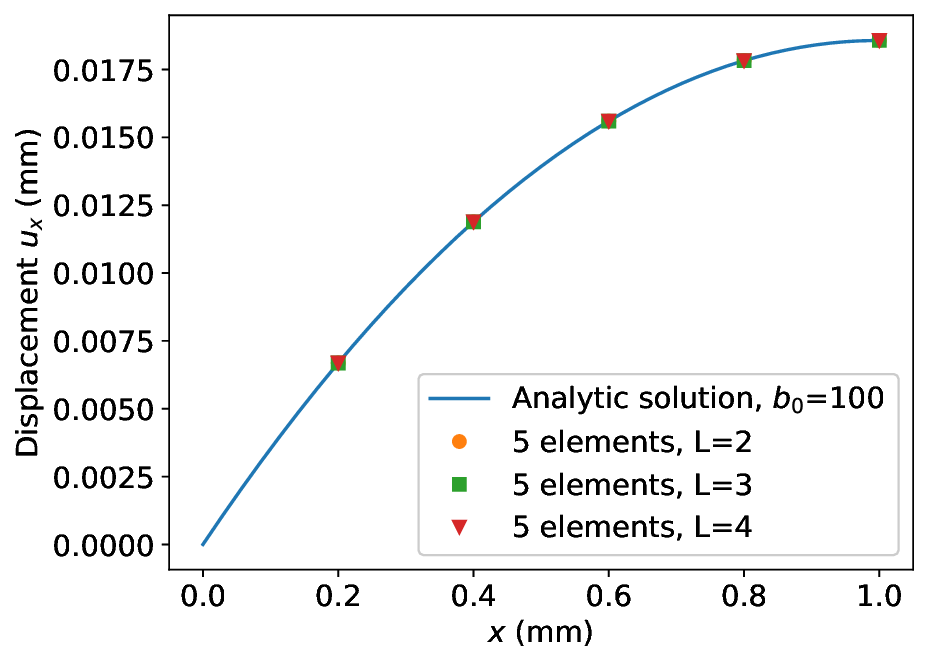}\\
		(a) 5 elements, convergence&(b) 5 elements, $u_x$\\
		\includegraphics[scale=0.5,valign=c]{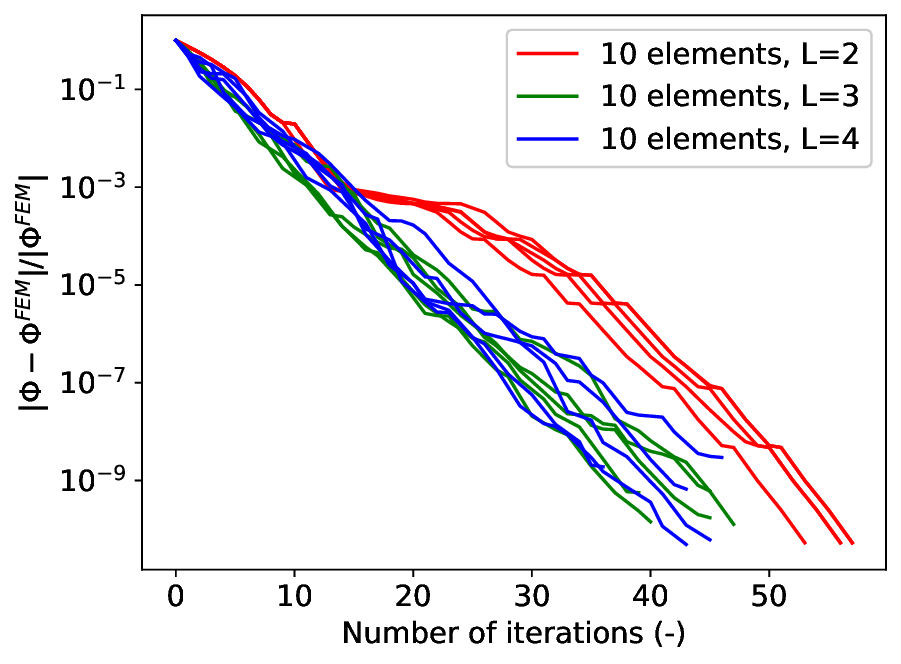}&\includegraphics[scale=0.5,valign=c]{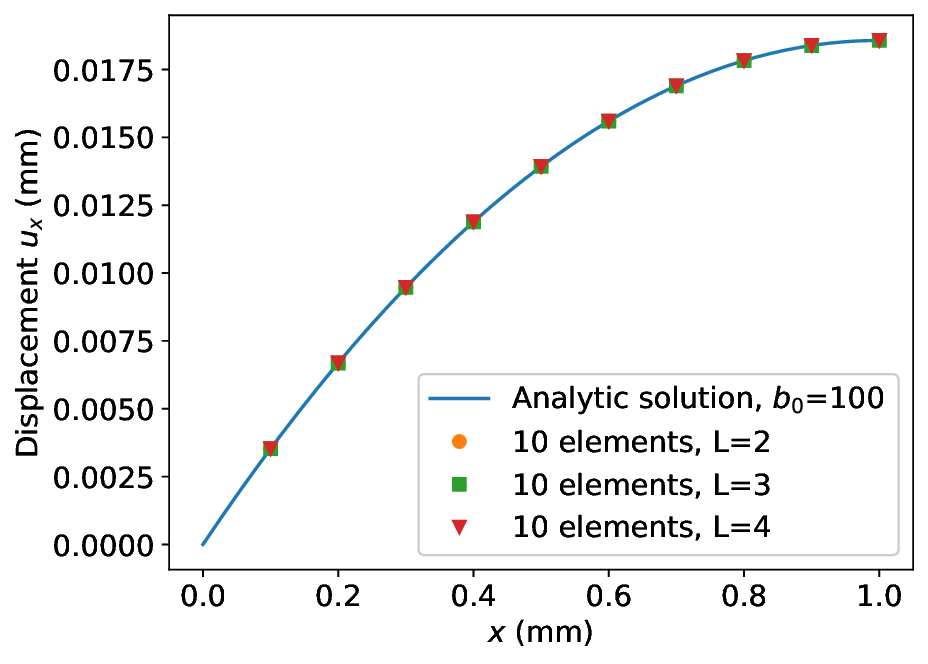}\\
		(c) 10 elements, convergence&(d) 10 elements, $u_x$\\
		\includegraphics[scale=0.5,valign=c]{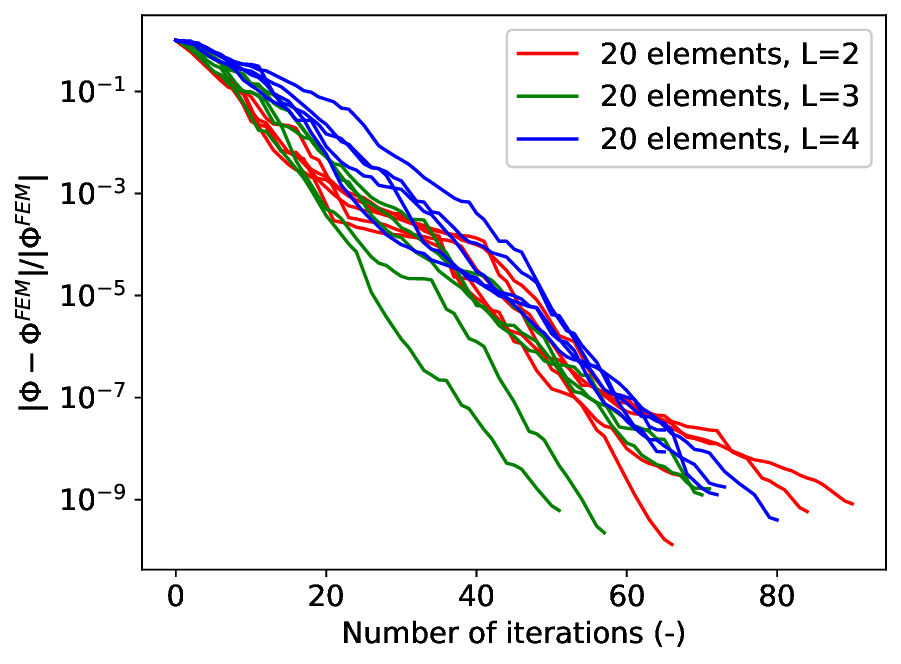}&\includegraphics[scale=0.5,valign=c]{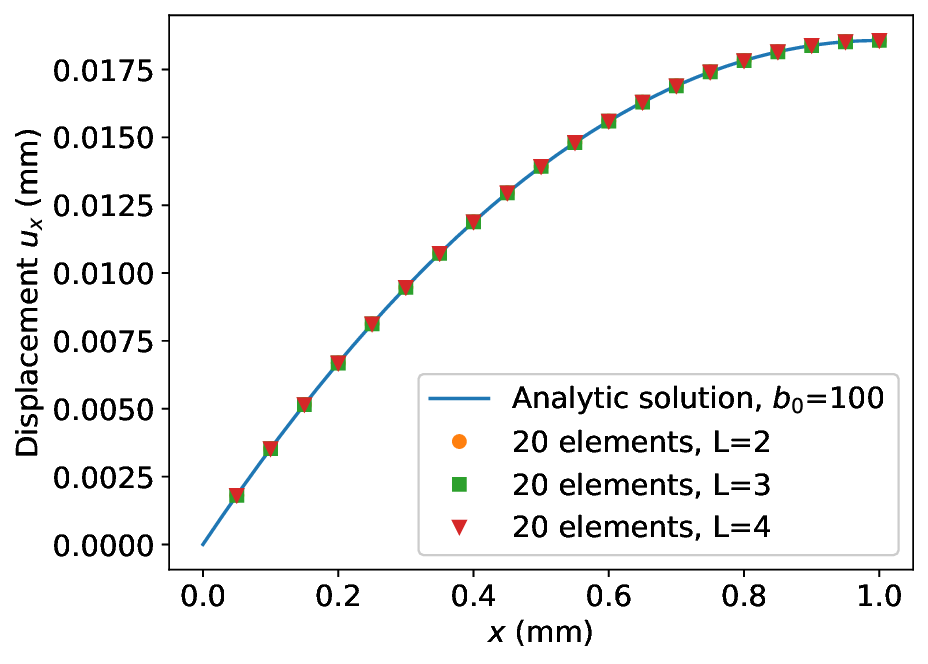}\\
		(e) 20 elements, convergence&(f) 20 elements, $u_x$
	\end{tabular}
	\caption{\firstreviewer{Uniaxial-strain tensile test with $b_0=100\,\text{MPa}\cdot\text{mm}^{-1}$: (a, c, e) convergence histories of 5 realisations for each value of $L$; and (b, d, f) displacement distribution over the bar predicted by the first realisation for each value of $L$. The default annealing time equal to $20 \mu s$ is considered.}}
	\label{fig:J2b100}
\end{figure}

\firstreviewer{The purely elastic case is first tested with $b_0=100$ MPa/mm. Different numbers of qubits, namely $L=$2, 3, and 4 qubits, per degree of freedom are successively considered for three successive discretisations of the bar into 5, 10, and 20 elements. Since the solution remains in the linear elastic domain, only a single minimisation problem of the total potential energy is performed. Each case is solved with five different realisations to study the randomness in the resolution.
    Figure \ref{fig:J2b100}(a,c,e) shows the convergence history of all realisations and Fig. \ref{fig:J2b100}(b,d,f) illustrates the displacement field distributions predicted by the first realisation of each case, which are also compared to the reported analytical results.
    Figure \ref{fig:J2b100}(a,c,e) shows that, for a given discretisation, the convergence histories are different for each realisation, although in the case of 5 elements and $L=2$, in which only a small number of qubits is required in the optimisation problem, the resolution exhibits no discrepancy. For all the studied cases, the accuracy keeps improving with the number of iterations, and relative errors as low as $1\times 10^{-9}$ --lower error can still be achieved by increasing the iterations number-- can be reached.}

\begin{figure}[!htb]
	\centering
	\begin{tabular}{cc}
		\includegraphics[scale=0.5,valign=c]{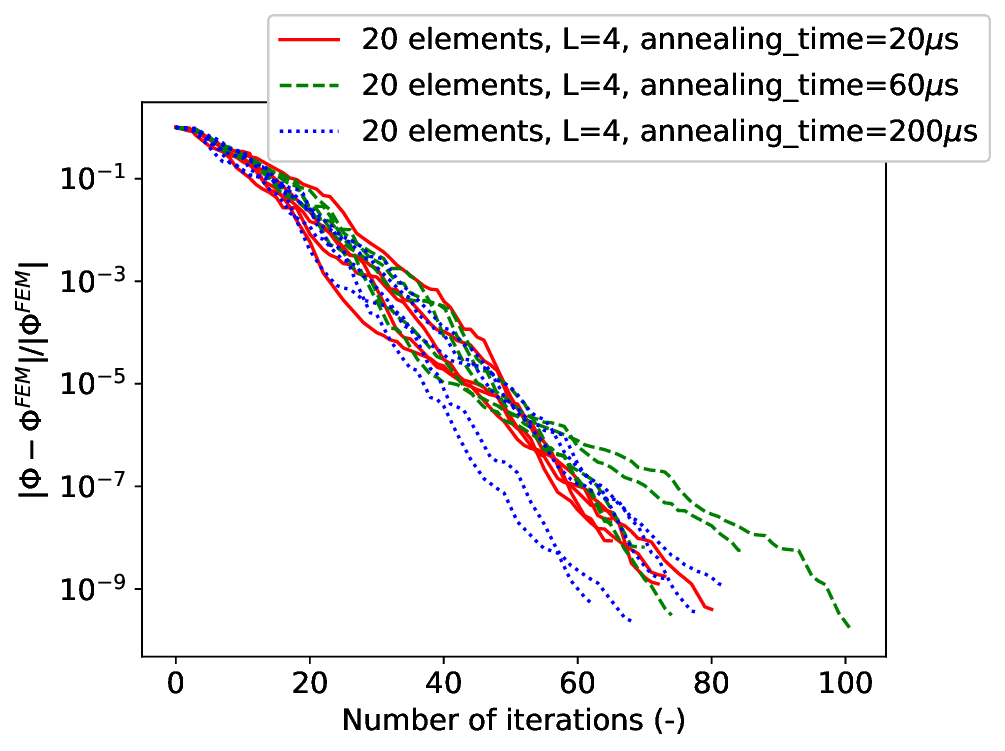}&\includegraphics[scale=0.5,valign=c]{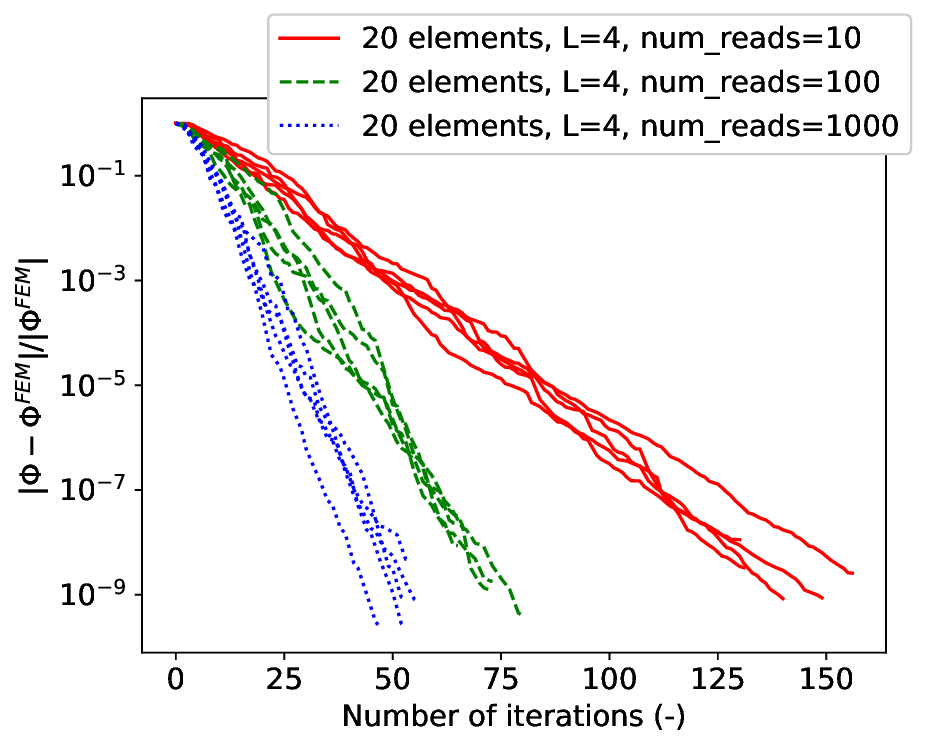}\\
		(a)&(b)
	\end{tabular}
	\caption{\secondreviewer{Uniaxial-strain tensile test with $b_0=100\,\text{MPa}\cdot\text{mm}^{-1}$: (a) effect of the annealing time (annealing\_time) for number\_reads=100; and (b) effect of the number of reads (number\_reads) for annealing\_time=20 $\mu$s on the convergence history.}}
	\label{fig:annealingTimeAndNumReads}
\end{figure}

\begin{figure}[!htb]
	\centering
	\includegraphics[scale=0.55,valign=c]{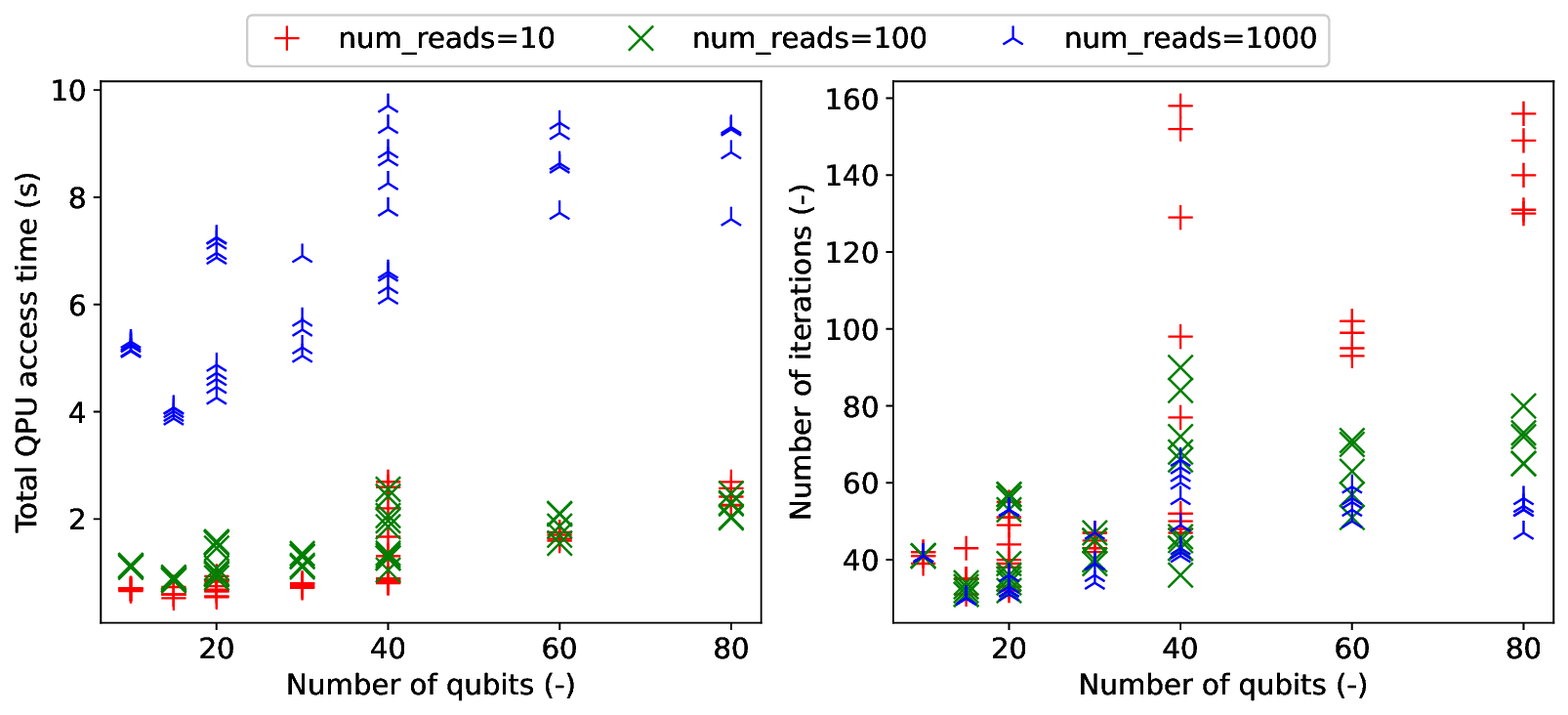}
	\caption{\secondreviewer{Uniaxial-strain tensile test with $b_0=100\,\text{MPa}\cdot\text{mm}^{-1}$: total QPU access time and number of iterations for each realisation in function of the number of qubits and number of reads  (number\_reads). The default annealing time equal to $20 \mu s$ is considered.}}
	\label{fig:timeAndNumIterations}
\end{figure}

\secondreviewer{From a practical point of view, the two most important user-defined parameters when using the DWave quantum annealers are the number of reads to evaluate the minimum of the objective function (num\_reads) and the annealing time (annealing\_time) as illustrated in Alg. \ref{alg:submitDwave}. Considering the elastic case with a 20-element discretisation and $L=4$,  Fig. \ref{fig:annealingTimeAndNumReads}(a) studies the effect of the annealing time (annealing\_time) successively taken as 20 $\mu$s (the default value), 60 $\mu$s, and 200 $\mu$s whilst the number of reads is kept constant and equal to 100. It is shown that, in this range, the annealing time does not significantly affect the convergence. Fig. \ref{fig:annealingTimeAndNumReads}(b) studies the influence of the number of reads (num\_reads) for a constant annealing time equal to 20 $\mu$s. The higher the number of reads, the faster the convergence: the current quantum hardware is sensitive to various factors, such as environmental noises, hardware imperfections, and pre- and post-processing errors, \emph{etc.}, so increasing the number of reads increases the probability of measuring the global minimum.

In order to assess the effect of the number of reads on the global computation time, the QPU access time for different realisations with different values of number of reads and different number of qubits $L$ per degree of freedom is reported in Fig. \ref{fig:timeAndNumIterations}(a). For a given number of qubits, a larger number of reads leads to a larger QPU access time, despite a lower total number of the iterations required to reach the same error, see Fig. \ref{fig:timeAndNumIterations}(b). We also observe the saturation of the QPU access time with the number of qubits for a given value of the number of reads, with a similar trend for the total number of iterations. We note that the field of quantum computing is rapidly evolving, and research is ongoing to improve the stability and scalability of quantum systems, so the QPU access time could be improved in the future.}

The elasto-plastic case is now considered with $b_0=400$ MPa/mm. The bar is discretised into 20 elements. Because of the existence of the plastic flow, the double-minimisation process detailed in Alg. \ref{alg:DoubleMinimisationFEM} is considered and each degree of freedom is discretised with three qubits.

\begin{figure}[!htb]
	\centering
	\begin{tabular}{cc}
		\includegraphics[scale=0.5,valign=c]{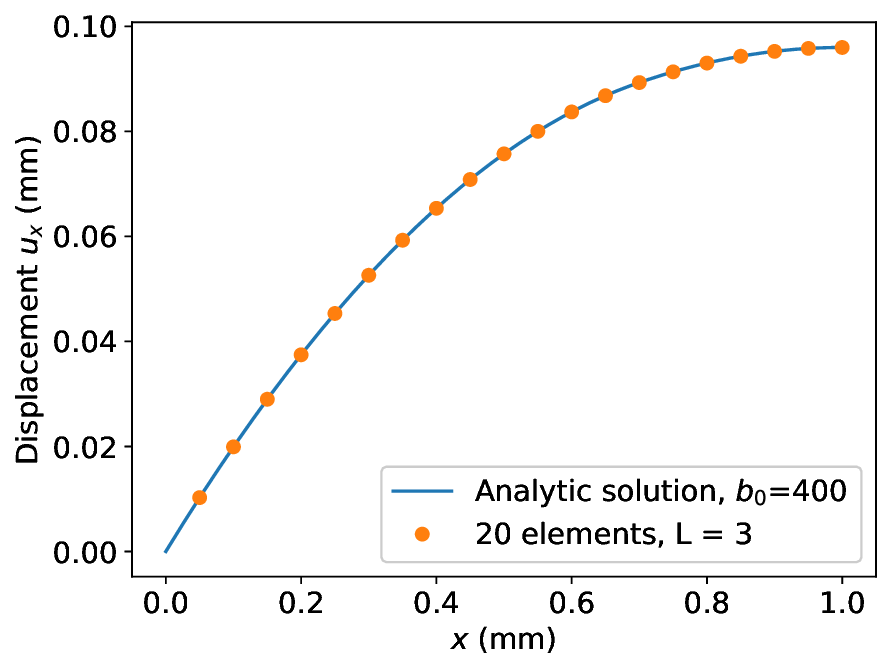}&\includegraphics[scale=0.5,valign=c]{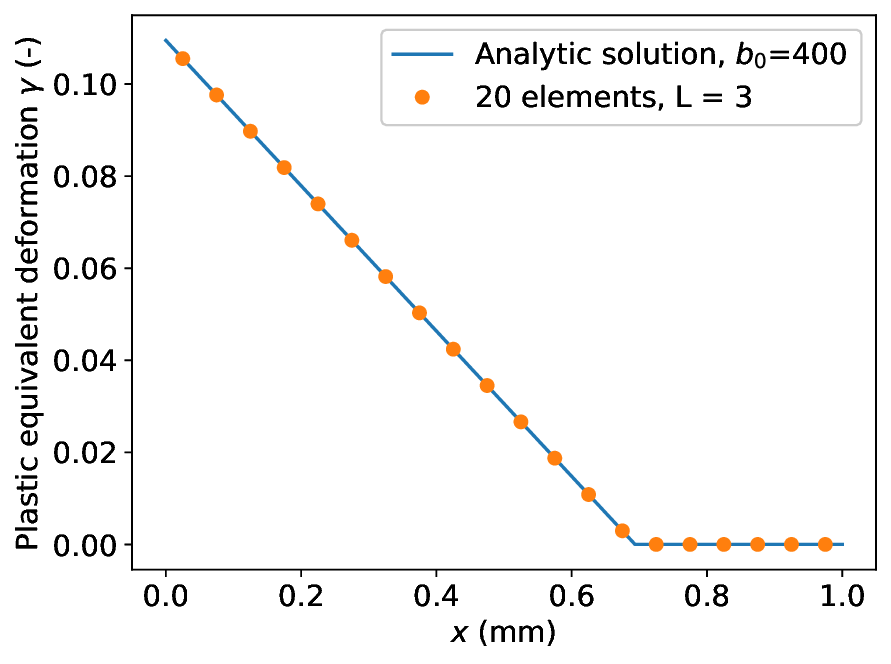}\\
		(a)&(b)
	\end{tabular}
	\caption{Uniaxial-strain tensile test with $b_0=400\,\text{MPa}\cdot\text{mm}^{-1}$: (a) displacement distribution; and (b) equivalent plastic strain distribution over the bar.}
	\label{fig:J2b400}
\end{figure}

\begin{figure}[!htb]
	\centering
	\begin{tabular}{cc}
		\includegraphics[scale=0.5,valign=c]{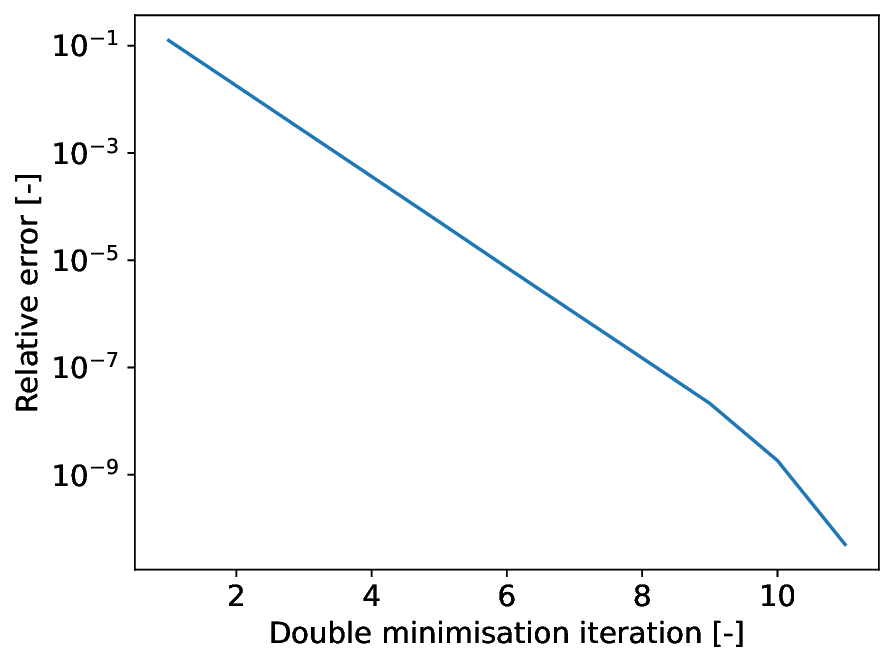}&\includegraphics[scale=0.5,valign=c]{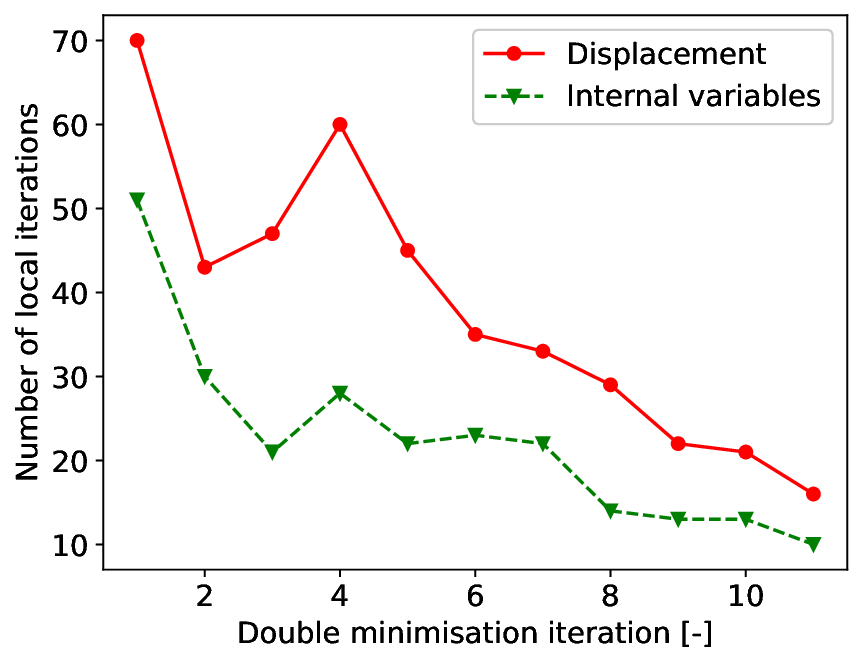}\\
		(a)&(b)
	\end{tabular}
	\caption{\firstreviewer{Uniaxial-strain tensile test with $b_0=400\,\text{MPa}\cdot\text{mm}^{-1}$: (a) convergence history in terms of the relative error \emph{vs.} the number of double-minimisation iterations and; (b) number of local iterations required at each double-minimisation iteration. The relative error is defined by  $|(\Phi_\mathbf{U}-\Phi_\mathbf{Q})/\Phi_{\mathbf{U}0}|$ where $\Phi_\mathbf{U}$ and $\Phi_{\mathbf{U}0}$ are respectively the objective function values after minimisation with respect to the displacement unknowns and the one at the first iteration, and $\Phi_\mathbf{Q}$ is the objective function value after minimisation with respect to the internal variable unknowns.}}
	\label{fig:OneDJ2errorIterationsAll}
\end{figure}
\begin{figure}[!htb]
	\centering
	\includegraphics[scale=0.5,valign=c]{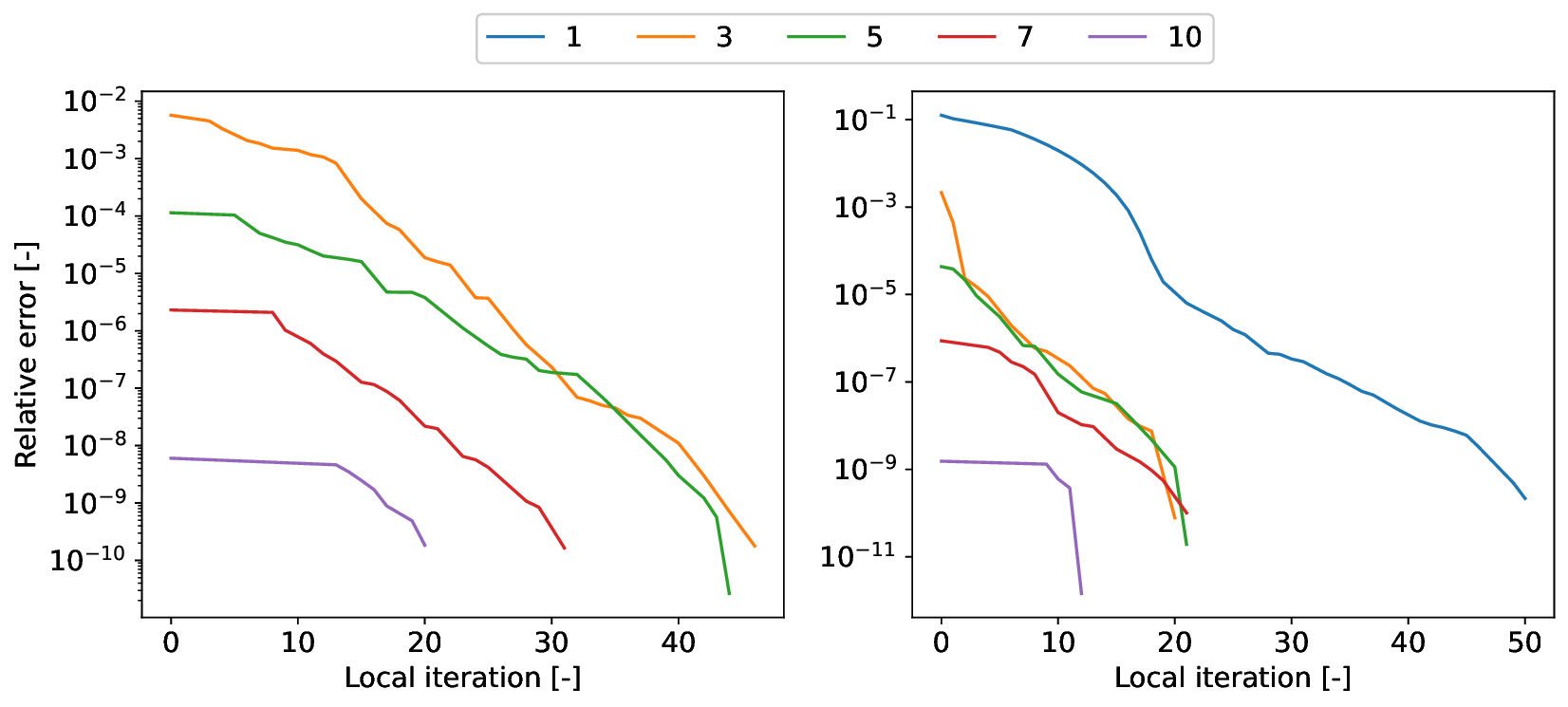}
	\caption{\firstreviewer{Uniaxial-strain tensile test with $b_0=400\,\text{MPa}\cdot\text{mm}^{-1}$: local iteration error convergence for the different double-minimisation iterations of (left) the displacement field optimisation; and (right) the internal variables field optimisation. The relative error is defined by $|(\Phi-\Phi_\text{-1})/\Phi_\text{0}|$ where $\Phi$ is the objective function value, and $\Phi_\text{0}$ and $\Phi_\text{-1}$ denote respectively the initial and last values of the objective function for each local minimisation.}}
	\label{fig:OneDJ2ConvergenceHistory}
\end{figure}

The obtained results in terms of the displacement distribution and of the equivalent plastic strain distribution are compared with the analytical results in Fig. \ref{fig:J2b400}, demonstrating that non-linear elasto-plastic problems can be solved with the quantum annealing hybrid approach. \firstreviewer{Figure \ref{fig:OneDJ2errorIterationsAll}(a) shows the convergence history of the error in terms of the number of double-minimisation iterations, \emph{i.e.} the number of iterations at line 3 of Alg. \ref{alg:DoubleMinimisationFEM}. For each double-minimisation iteration, the two local minimisation problems in terms of the displacement field and of the internal variables field, respectively lines 4 and 5 of Alg. \ref{alg:DoubleMinimisationFEM}, are solved through local iterations during the QA-SQP process following Alg. \ref{alg:QA}. The number of local iterations required for each local minimisation problem to reach a given accuracy of $1\times 10^{-9}$ is reported in Fig. \ref{fig:OneDJ2errorIterationsAll}(b), in which it can be seen that the number of local iterations decreases with the convergence of the double-minimisation problem. The convergence histories of the local displacement field optimisation problem and of the internal variables field optimisation at different double-minimisation iteration steps are shown in Fig. \ref{fig:OneDJ2ConvergenceHistory}, showing that relatively low errors of $1\times 10^{-9}$ can be reached during the iterative process.}

\subsection{Elasto-plastic plate under tensile loading}\label{subsec:2DTest}

\begin{figure}[!htb]
	\centering
	\includegraphics[scale=0.5,valign=c]{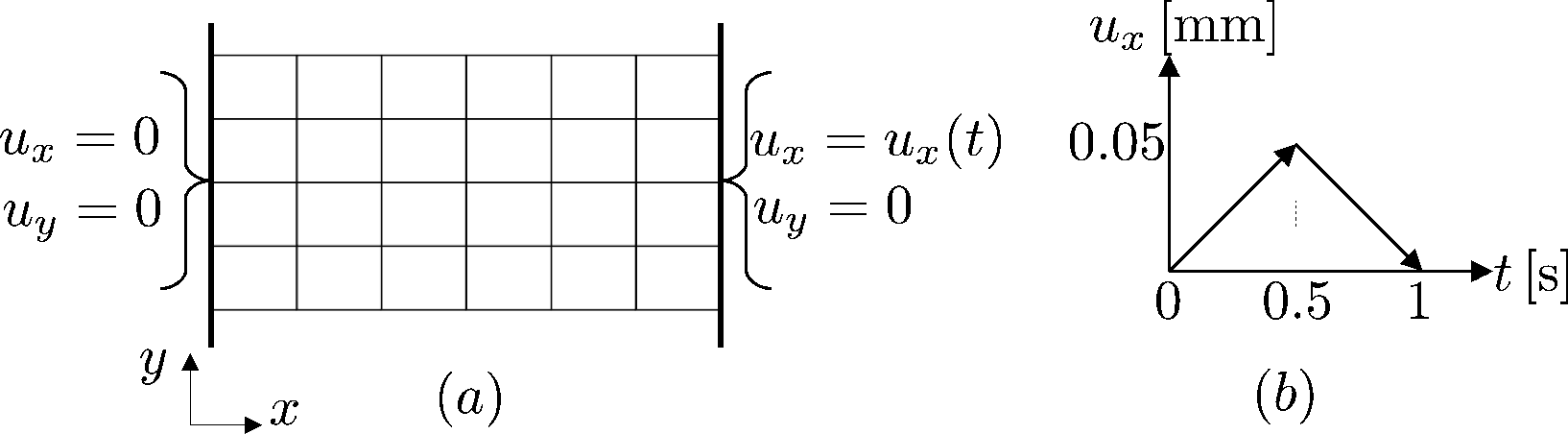}
	\caption{Plate under cyclic loading: (a) the finite element mesh of the $1$mm$\times0.5$mm-plate under prescribed displacement boundary conditions; and (b) the cyclic displacement path applied to the right edge of the plate.}
	\label{fig:2DTest}
\end{figure}

In this section, we consider the proposed hybrid framework to study the cyclic loading of a two-dimensional plate under plane-strain condition, in order to demonstrate the capability of the framework to account for history-dependency.
The finite element mesh and the boundary conditions are sketched in Fig. \ref{fig:2DTest}(a).
The cyclic prescribed loading displacement is applied on the right edge of the plate as shown in Fig. \ref{fig:2DTest}(b). The material of the plate obeys a $J_2$-elasto-plastic model with a Swift isotropic hardening law as reported in Tab. \ref{tab:materialParameters2}. The current solution is compared to the one obtained with a conventional finite element method (FEM).

\begin{table}[htb]
	\centering
	\caption{Material properties considered in the plate under cyclic loading.}
	\label{tab:materialParameters2}
	\begin{tabular}{|l|c|c|c|}
		\hline
		Parameter&Notation/Expression&Value/Expression&Unit\\
		\hline
		Young's modulus&$E$&20&GPa\\
		\hline
		Poisson's ratio&$\nu$&0.3&-\\
		\hline
		Initial yield stress&$\sigma_y^0$&150&MPa\\
		\hline
		Isotropic hardening&$R$&$\sigma_y^0\left[\left(1+\cfrac{\gamma}{0.05}\right)^{0.1}-1\right]$&MPa\\
		\hline
	\end{tabular}
\end{table}

\begin{figure}[!htb]
	\centering
	\includegraphics[scale=0.5,valign=c]{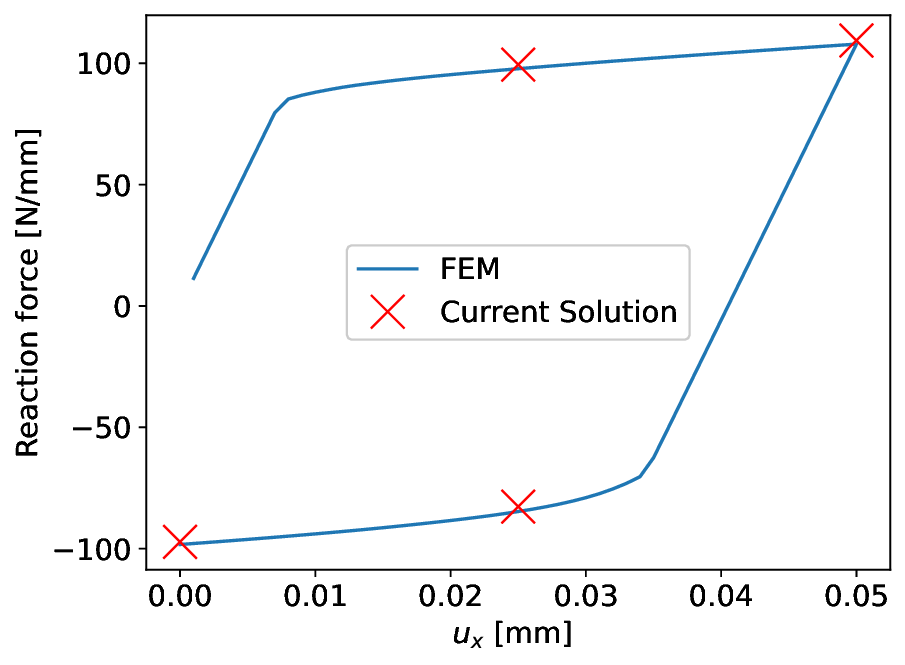}
	\caption{Plate under cyclic loading: comparison between the current framework and the classical FEM in terms of the reaction force on the left boundary of the plate \emph{vs.} the prescribed displacement.}
	\label{fig:2DTestForceDisp}
\end{figure}

\begin{figure}[!htb]
	\centering
	\includegraphics[scale=0.5,valign=c]{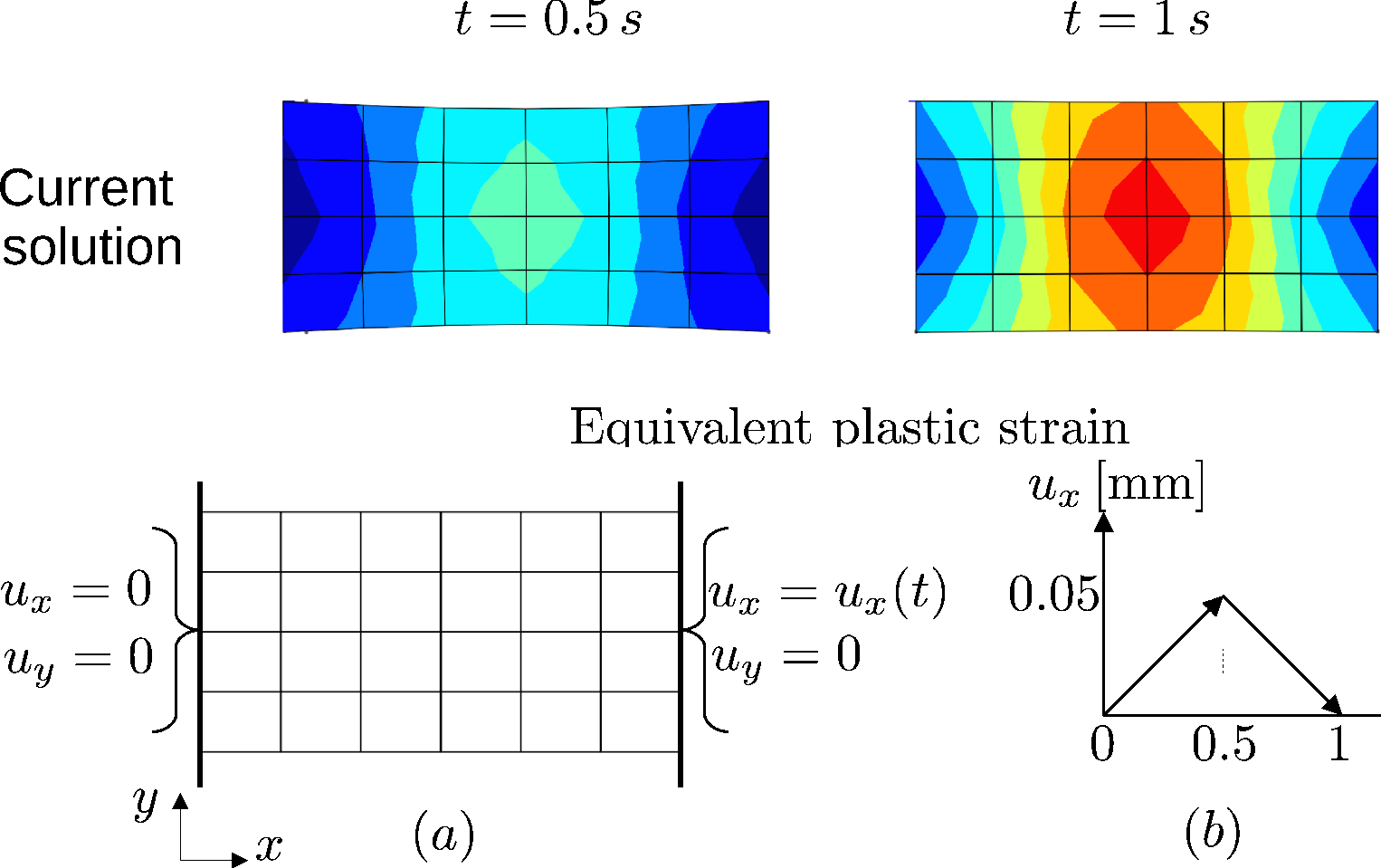}
	\caption{Plate under cyclic loading: comparison between the current framework and the classical FEM in terms of the distribution of the equivalent plastic strain at $t=0.5$ s and at $t=1$ s.}
	\label{fig:2DTestEPS}
\end{figure}

The current double-minimisation framework assisted by quantum annealing is conducted at four different times, \emph{i.e.} four increments are considered to solve the problem. \firstreviewer{Since the double-minimisation process can handle large increments (this is due to the fact that the elastic and plastic energies are correctly introduced), the whole simulation can be conducted using the developed QA-SQP in two increments only. Four increments are however considered in order to have intermediate solutions that can be compared with the classical finite element results.} The considered times are $t=0.25$s, $t=0.5$s, $t=0.75$s and $t=1$s. Each degree of freedom is discretised with three qubits. Figures \ref{fig:2DTestForceDisp} and \ref{fig:2DTestEPS} compare the current framework with a classical FEM, respectively in terms of the reaction force on the left boundary of the plate \emph{vs.} the prescribed displacement and of the distribution of the equivalent plastic strain at two configurations corresponding to $t=0.5$s and $t=1$s, see Fig. \ref{fig:2DTest}(b), demonstrating that the quantum annealing hybrid framework is able to model history-dependent behaviours. 

\begin{figure}[!htb]
	\centering
	\begin{tabular}{cc}
		\includegraphics[scale=0.5,valign=c]{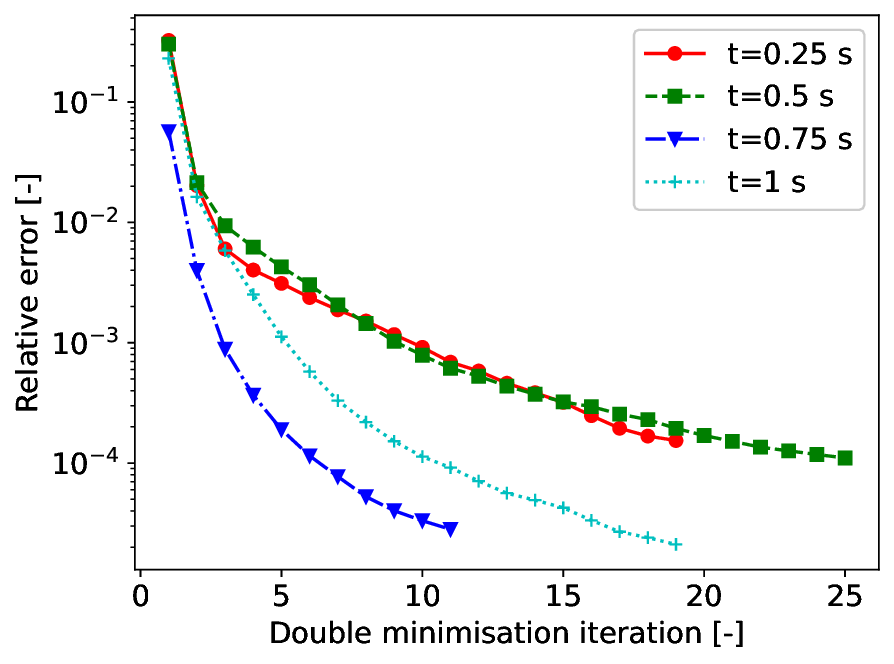}&\includegraphics[scale=0.5,valign=c]{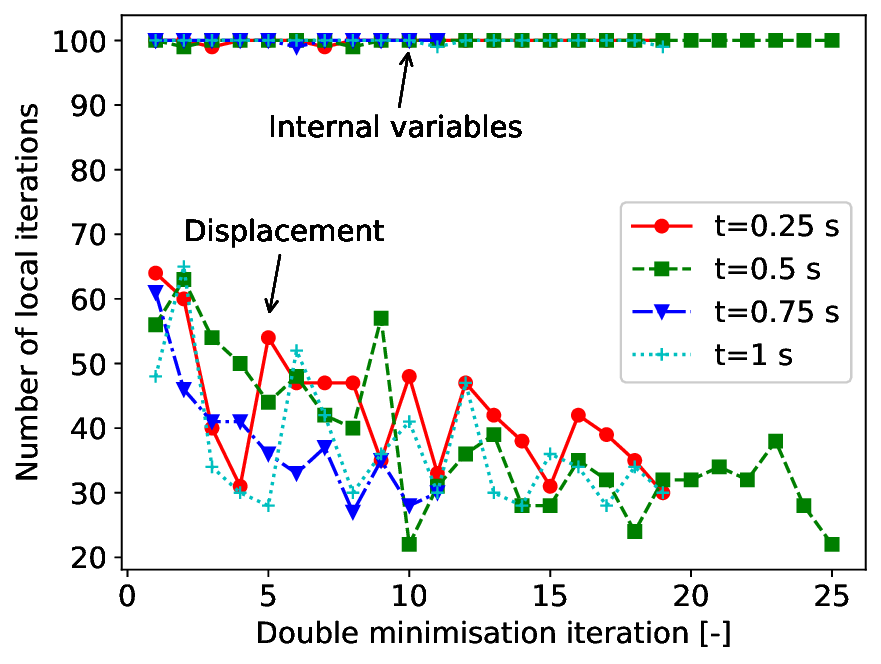}\\
		(a)&(b)
	\end{tabular}
	\caption{\firstreviewer{Plate under cyclic loading: (a) convergence history in terms of the relative error \emph{vs.} the number of double-minimisation iterations and; (b) number of local iterations required at each double-minimisation iteration. The relative error is defined by  $|(\Phi_\mathbf{U}-\Phi_\mathbf{Q})/\Phi_{\mathbf{U}0}|$ where $\Phi_\mathbf{U}$ and $\Phi_{\mathbf{U}0}$ are respectively the objective function values after minimisation with respect to the displacement unknowns and the one at the first iteration, and $\Phi_\mathbf{Q}$ is the objective function value after minimisation with respect to the internal variable unknowns. }}
	\label{fig:errorIterationsAll}
\end{figure}

\begin{figure}[!htb]
	\centering
	\includegraphics[scale=0.5,valign=c]{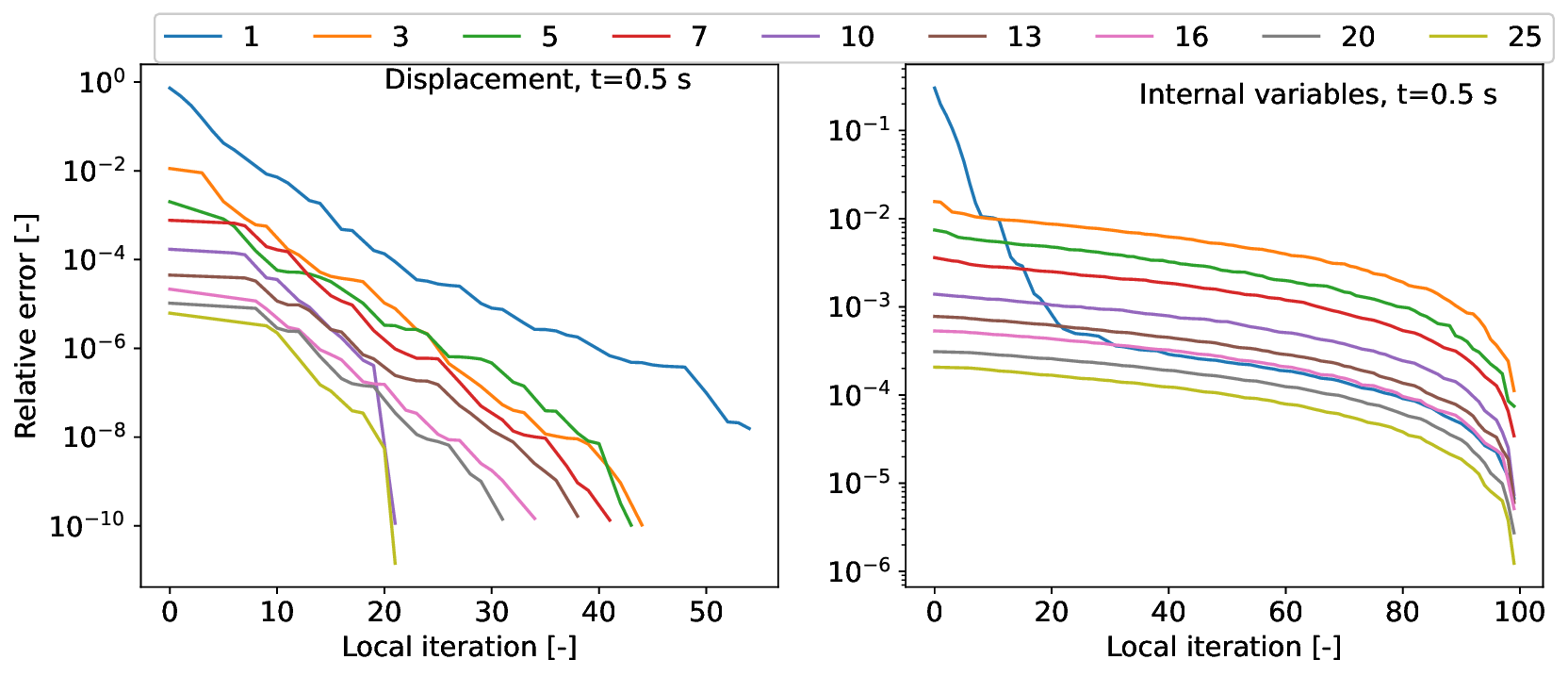}
	\caption{\firstreviewer{Plate under cyclic loading:local iteration error convergence for the different double-minimisation iterations of (left) the displacement field optimisation; and (right) the internal variables field optimisation. The relative error is defined by $|(\Phi-\Phi_\text{-1})/\Phi_\text{0}|$ where $\Phi$ is the objective function value, and $\Phi_\text{0}$ and $\Phi_\text{-1}$ denote respectively the initial and last values of the objective function for each local minimisation.}}
	\label{fig:convergenceHistoryStep2}
\end{figure}

\firstreviewer{Figure \ref{fig:errorIterationsAll}(a) shows the error evolution with the number of double-minimisation iterations, \emph{i.e.} the number of iterations at line 3 of Alg. \ref{alg:DoubleMinimisationFEM}, at the four successive configuration resolutions. The numbers of local iterations required for each local minimisation problem, \emph{i.e.} the QA-SQP resolution following Alg. \ref{alg:QA} at lines 4 and 5 of Alg. \ref{alg:DoubleMinimisationFEM}, are reported in Fig. \ref{fig:errorIterationsAll}(b) for each of the double-minimisation iteration. The number of iterations for the minimisation problem with respect to the internal variables field is limited to 100 in order to limit the queue waiting time. It is however noted that the double-minimisation strategy sketched in Alg. \ref{alg:DoubleMinimisationFEM} does not require each local minimisation to be exact in order to reach the convergence of the double-minimisation problem, but only a solution improvement after each local minimisation. The convergence histories of the local displacement field optimisation problem and of the internal variables field optimisation at $t=0.5$ s are shown in Fig. \ref{fig:convergenceHistoryStep2} for different double-minimisation iteration steps, demonstrating the decrease of the objective functions after each minimisation problem resolution.}

\section{Conclusions} \label{sec:conclusions}

Quantum annealing has the potential to enhance finite element simulations in computational solid mechanics to study larger problem sizes, thus offering the prospect of more efficient and accurate solutions when analysing complex mechanical problems. Our study explores the possibility of quantum annealing techniques to be used as a solver in the context of the finite element method in order to tackle, in the future, computationally demanding problems, thus enabling researchers and engineers to explore new avenues in the design, analysis, and optimisation of mechanical structures and systems.

Towards this end, the weak form was reformulated as a multi-optimisation problem solving both the material constitutive models and the structure balance equations at the structural level of the finite element problem. 
After transforming the non-linear optimisation problems into series of quadratic ones, and after binarisation of the continuous degrees of freedom, the minimisation steps were shown to be compatible with the quantum annealer.
As a result, the system resolution steps, included the ones related to the non-linear system of equations of the material model, are conducted on the quantum annealer, whilst the classical computer is used for the finite element assembly.
The hybrid formulation was then applied on elasto-plastic 1D and 2D problems, in which is was shown that, on the one hand, error lower than $1 \time 10^{-9}$ can be achieved by considering 4 qubits to binarise a degree of freedom and that, on the other hand, that history-dependent behaviours can be modelled.

\secondreviewer{Although simple constitutive material models such as the presented J2-plasticity model do not present difficulty in their resolution with classical computers, this is not true for more complex material models, \emph{e.g.} such as for finite-strain visco-elasticity visco-plasticity with pressure dependent yield surface, which involve several internal variables and for which deriving and evaluating the material operator require many algorithmic operations that slow down the resolution process. It is expected that with the development of quantum computers, the developed double-minimisation QA-SQP strategy, in which all the resolution steps are conducted on the quantum annealer whilst only assembly steps remain on the classical hardware, will present a clear advantage in the context of complex finite element simulations since quantum annealing is particularly efficient in evaluating the minimum point of a multi-variate quadratic function. Additionally, the double-minimisation strategy offers two other advantages: (i) as shown in Section \ref{subsec:doubleMinFE}, the minimisation problem in terms of the displacement field considers an elastic behaviour at a given internal variables state, which makes the material operator either easier to be evaluated or constant when defining the quadratic functional, and which is beneficial in terms of computational efficiency, and (ii) the minimisation problem in terms of the internal variables is simply the combination of independent constrained minimisation problems at each integration point since the displacement field is fixed; it thus does not require all the qubits to be connected and can therefore exploit the parallelism advantage of quantum computers.}

We note that currently the size of the problem that can be solved is limited because of the current limitation in terms of available physical qubits number and of their interactions. Nevertheless, it is expected that with the progress in quantum computer hardware, problems of size larger that what can currently be handled with classical computers will become solvable.

\section*{Acknowledgments}
V.-D.N acknowledges the support of the Fonds National de la Recherche (F.R.S.-FNRS, Belgium).

F.R. acknowledges the support of the Fonds National de la Recherche (F.R.S.-FNRS, Belgium), \#T0205.20.

This work is partially supported by a ``Strategic Opportunity'' grant from the University of Liege.

\section*{Data availability}
The code and raw/processed data required to reproduce these findings is available on \cite{DATAQC} under the Creative Commons Attribution 4.0 International (CC BY 4.0) licence


\appendix
\section{Finite element approximation}\label{sec:femApp}

The domain $V$ is discretised into a finite element mesh of $N_\text{ele}$ elements, \emph{i.e.}
\begin{eqnarray}
	V = \bigcup_{e=0}^{N_\text{ele}-1} V^e\,.
\end{eqnarray}
The Neumann boundary $\partial_N V$ is discretised accordingly into $N_\text{ele}^\text{neu}$ elements with a lower dimension as
\begin{eqnarray}
	\partial_N V = \bigcup_{e=N_\text{ele}}^{N_\text{ele}+N_\text{ele}^\text{neu}-1}\Gamma^e\,.
\end{eqnarray} 
On an arbitrary element $V^e$ (or $\Gamma^e$), we approximate the displacement field $\vb{u}$ and internal variables field $\vb{q}$ under the following interpolants\footnote{For ease of readability, the notation $\bullet_a^e$ refers to the quantity $\bullet$ on the node $a$ of element $e$ whilst $\bullet^{e,\eta}$ refers to the quantity $\bullet$ on the quadrature point $\eta$ of element e.}
\begin{eqnarray} 
	\vb{u}\left(\vb{x}\right) &=& \sum_{a=0}^{n_e} {N}^e_a\left(\vb{x}\right) \vb{u}_a^e \,,\text{ and} \label{eq:dispAppro}\\
	\vb{q}\left(\vb{x}\right) &=& \sum_{\eta=0}^{m_e} {M}^{e,\eta}\left(\vb{x}\right) \vb{q}^{e,\eta}\,, \label{eq:qApprox}
\end{eqnarray}
where $N^e_a$ and $M^e_\eta$ are respectively the displacement and internal variables shape functions, $n_e$ and $m_e$ are respectively the number of nodes and quadrature points on element $e$, and where $\vb{u}_a^e$ and $\vb{q}^{e,\eta}$ are the displacement and internal variables at the sampled points. On the one hand, the displacement field needs to be continuous in $V$, \emph{i.e.} $\vb{u}\in C_0\left(V\right)$. This can be easily achieved using the isoparametric finite elements with Lagrange shape functions, in which the Kronecker delta property is satisfied at all nodes of the elements, \emph{i.e.}
\begin{eqnarray}
	{N}^e_a\left(\vb{x}_b^e\right) = \delta_{ab} \,,\; \forall e, a, b\,,
\end{eqnarray}
where $\vb{x}_b^e$ for $b=0,\ldots, n_e-1$ are the coordinates of the nodes of element $e$. By contrast, since the internal variables field needs to be bounded, \emph{i.e.} $\vb{q} \in L^\infty\left(V\right)$, the shape functions are assumed to satisfy the Kronecker delta property at quadrature points as
\begin{eqnarray}
	{M}^{e,\eta}\left(\bm{\Xi}^{e,\zeta}\right) =\delta_{\eta\zeta} \,,\; \forall e, \eta, \zeta\,,
\end{eqnarray}
where $\bm{\Xi}^{e,\zeta}$ for $\zeta=0, \ldots, m_e-1$ are the quadrature points in element $e$.
We note that the internal variables field $\vb{q}$ is only used at the quadrature points, so that the interpolation (\ref{eq:qApprox}) is licit.

Clearly, using finite element interpolants (\ref{eq:dispAppro}, \ref{eq:qApprox}), the displacement unknowns vector (denoted by $\vb{U}$) includes all displacements at the nodes of the mesh except the ones at the nodes on the Dirichlet boundary $\partial_D V$ where the displacement field is prescribed, whilst the internal variable unknowns vector (denoted by $\vb{Q}$) includes all the internal variables at the quadrature points of the elements.

Eq. (\ref{eq:dispAppro}) evaluated at the quadrature points can be organised in a matrix-vector operation as
\begin{eqnarray}\label{eq:dispApproMatrix}
	\vb{u}^{e,\eta} = \vb{u}\left(\bm{\Xi}^{e,\eta}\right) = \vb{N}^{e,\eta} \vb{U}^e \,,
\end{eqnarray}
where $\vb{N}^{e,\eta}$ is the elementary shape function matrix evaluated at the quadrature point $\bm{\Xi}^{e,\eta}$ and $\vb{U}^e$ is the elementary unknowns vector gathering all elementary nodal displacements. Additionally, we introduce the operator $\mqty[\bullet]$ to convert, respectively, a second-order tensor and a fourth-order tensor to a vector and a square matrix as follows
\begin{eqnarray}
	\mqty[\bullet]_{id+j} = \vb{\bullet}_{ij} \text{ and } \mqty[\bullet]_{id+j,kd+l} = \vb{\bullet}_{ijkl} \,,
\end{eqnarray}
where $d$ is the dimension. As a result, Eq. (\ref{eq:dispApproMatrix}) yields
\begin{eqnarray}\label{eq:FEMApp}
	\mqty[\bm{\varepsilon}^{e,\eta}] =  \mqty[\eval{\grad\otimes^s \vb{u}}_{\bm{\Xi}^{e,\eta}}] = \vb{B}^{e,\eta}\vb{U}^e\,,
\end{eqnarray}
where $\vb{B}^{e,\eta}$ is the elementary matrix of the gradient of the shape functions associated with the displacement vector field at the quadrature point $\bm{\Xi}^{e,\eta}$.

The energy function (\ref{eq:energyVar}) is estimated at a quadrature point, leading to 
\begin{eqnarray}
	\Delta \Phi_{n+1} &=&  \Delta W^\text{int}_{n+1} - \Delta W^\text{ext}_{n+1}\,, \text{ with } \label{eq:energyVar2} \\
	\Delta W^\text{int}_{n+1} &=& \sum_{e=0}^{N_\text{ele}-1}\sum_{\eta=0}^{m_e-1} \omega^{e,\eta} \Delta \mathcal{E}_{n+1}\left(\Delta \bm{\varepsilon}_{n+1}^{e,\eta},\Delta\vb{q}_{n+1}^{e,\eta} \right) \,,\text{ and} \label{eq:WintIncr}\\
	\Delta W^\text{ext}_{n+1} &=& \sum_{e=0}^{N_\text{ele}-1}\sum_{\eta=0}^{m_e-1} \omega^{e,\eta} \vb{b}_0^{e,\eta}\cdot\Delta\vb{u}^{e,\eta}_{n+1}  + \sum_{e=N_\text{ele}}^{N_\text{ele}+N_\text{ele}^\text{neu}-1}\sum_{\eta=0}^{m_e-1} \omega^{e,\eta}\vb{t}_0^{e,\eta} \cdot\Delta\vb{u}^{e,\eta}_{n+1}\,, \label{eq:WextIncr}
\end{eqnarray}
where $\bullet^{e,\eta}$ denotes a quantity $\bullet$ at the quadrature point $\eta$ of element $e$, and $\omega^{e,\eta}\,,\;\forall e, \eta$, denotes the weights of the corresponding quadrature points $\bm{\Xi}^{e,\eta}$. Using Eq. (\ref{eq:FEMApp}), Eqs. (\ref{eq:WintIncr}, \ref{eq:WextIncr}) become
\begin{eqnarray}
	\Delta W^\text{int}_{n+1} &=& \sum_{e=0}^{N_\text{ele}-1}\sum_{\eta=0}^{m_e-1} \omega^{e,\eta} \Delta \mathcal{E}_{n+1}\left( \vb{B}^{e,\eta}\Delta\vb{U}^e_{n+1},\Delta\vb{q}_{n+1}^{e,\eta} \right) \,,\text{ and} \label{eq:WintIncr2} \\	
	\Delta W^\text{ext}_{n+1} &=& \Delta \vb{U}_{n+1}^T \mathbf{f}^\text{ext}_{n+1}\,,
\end{eqnarray}
where $\mathbf{f}^\text{ext}_{n+1}$ is the external force vector
\begin{eqnarray}\label{eq:Fext}
	\mathbf{f}^\text{ext}_{n+1} = \assembleOperator_{V^e \subset V} \sum_{\eta=0}^{m_e-1} \omega^{e,\eta} \left(\vb{N}^{e,\eta}\right)^T \vb{b}_0^{e,\eta}  + \assembleOperator_{\Gamma^e \subset \partial_N V} \sum_{\eta=0}^{m_e-1}  \omega^{e,\eta} \left(\vb{N}^{e,\eta}\right)^T \vb{t}_0^{e,\eta}\,.
\end{eqnarray}
with $\assembleOperator$ being used to indicate the assembling process.

\section{Analytical solution of an elasto-plastic bar under uniaxial strain test} \label{app:UniaxialTestAnalytical}

The plastic deformation is deviatoric, implying the following plastic deformation tensor
\begin{eqnarray}
	{\bm{\varepsilon}}^{\text{p}} =  \mqty[\varepsilon^{\text{p}}_{xx}&0&0\\0&-\cfrac{1}{2}\varepsilon^{\text{p}}_{xx}&0\\0&0&-\cfrac{1}{2}\varepsilon^{\text{p}}_{xx}]\,,
\end{eqnarray}
where $\varepsilon^{\text{p}}_{xx} > 0$ as the result of the monotonic tensile loading. The equivalent plastic strain is given as
\begin{eqnarray}
	\dot{\gamma} = \sqrt{\frac{2}{3} \dot{\bm{\varepsilon}}^{\text{p}}: \dot{\bm{\varepsilon}}^{\text{p}}} = \dot{\varepsilon}^{\text{p}}_{xx}\,,
\end{eqnarray}
implying $\gamma={\varepsilon}^{\text{p}}_{xx}$. As a result, the Hooke's law can be rewritten as
\begin{eqnarray}
	\bm{\sigma} = K\varepsilon_{xx} \vb{I} + 2\mu \mqty[\cfrac{2}{3}\varepsilon_{xx}-\gamma&0&0\\0&\cfrac{\gamma}{2}-\cfrac{\varepsilon_{xx}}{3}&0\\0&0&\cfrac{\gamma}{2}-\cfrac{\varepsilon_{xx}}{3}]\,,
\end{eqnarray}
where $K$ and $\mu$ are the shear and bulk moduli, respectively. The last equation leads to
\begin{eqnarray}
	\sigma_{xx} &=& \left(K+ \frac{4}{3}\mu\right){\varepsilon}_{xx} - 2\mu \gamma \,,\text{ and } \\
	\sigma_{\text{eq}} &=& \sqrt{\cfrac{3}{2}\dev{\bm{\sigma}}:\dev{\bm{\sigma}}} = 3\mu \left(\cfrac{2}{3}\varepsilon_{xx}-\gamma\right)\,.
\end{eqnarray}
Equations (\ref{eq:1Dbalance}) lead to the solution of $\sigma_{xx}$ as
\begin{eqnarray}
	\sigma_{xx} = b_0\left(l- x\right)\,.
\end{eqnarray}
The Kuhn-Tucker condition reads
\begin{eqnarray}\label{eq:KT2}
	\sigma_{\text{eq}} - \sigma_y^0 - H\gamma \leq 0 \,,\gamma \geq 0\,, \text{ and } \left[\sigma_{\text{eq}} - \sigma_y^0 - H\gamma\right]\gamma = 0\,,
\end{eqnarray}
where $\sigma_y^0$ is the onset of plasticity and $H$ is the linear hardening modulus. 

Finally, one has the following equations for $\varepsilon_{xx}$, $\gamma$, and $u_x$:
\begin{eqnarray}
	\begin{cases}
		&\left(K+ \frac{4}{3}\mu\right){\varepsilon}_{xx} - 2\mu \gamma = b_0\left( l- x\right) \,,\\
		&3\mu \left(\cfrac{2}{3}\varepsilon_{xx}-\gamma\right) - \left(\sigma_y^0 + H\gamma\right) \leq 0 \,,
		\gamma \geq 0 \,, \\
		&\left[3\mu \left(\cfrac{2}{3}\varepsilon_{xx}-\gamma\right) - \left(\sigma_y^0 + H\gamma\right)\right]\gamma = 0\,, \text{ and }\\
		& u_x = \int_{0}^x \varepsilon_{xx}\,dx\,.
	\end{cases}
\end{eqnarray}
The system of equations above yields the following solution
\begin{eqnarray}
	&&\text{ if } x \geq x_c \,,	\begin{cases}
		\varepsilon_{xx} &= \cfrac{b_0\left(l- x\right)}{K+ \cfrac{4}{3}\mu}  \\
		\gamma &= 0 \\
		u_x& = \begin{cases} \cfrac{b_0}{K+ \cfrac{4}{3}\mu}\left(lx - \cfrac{x^2}{2} \right) &\text{ if } x_c < 0 \\
			u_c + \cfrac{b_0}{K+ \cfrac{4}{3}\mu}\left[l\left(x -x_c\right) - \cfrac{x^2-x_c^2}{2} \right] &\text{ if } x_c \geq 0 
		\end{cases}
	\end{cases}  \,,\text{ and } \nonumber \\
	&&\text{ if } x < x_c \,,	\begin{cases}
		\varepsilon_{xx} &= \cfrac{1}{K+ \cfrac{4}{3}\mu - \cfrac{4\mu^2}{3\mu+H}} \left[b_0\left( l- x\right) - \cfrac{2\mu \sigma_y^0}{3\mu+H}\right] \\
		\gamma &= \cfrac{2\mu }{3\mu + H} \varepsilon_{xx} -\cfrac{ \sigma_y^0}{3\mu + H} \\
		u_x &= \cfrac{1}{K+ \frac{4}{3}\mu - \cfrac{4\mu^2}{3\mu+H}} \left[b_0\left(l x- \cfrac{x^2}{2}\right) - \cfrac{2\mu \sigma_y^0 x}{3\mu+H}\right]
	\end{cases} 
\end{eqnarray}
where 
\begin{eqnarray}
	x_c &=& l- \left(K+ \frac{4}{3}\mu\right)\frac{\sigma_y^0}{2\mu b_0}\,,\text{ and }\\
	u_c &=& \cfrac{1}{K+ \frac{4}{3}\mu - \cfrac{4\mu^2}{3\mu+H}} \left[b_0\left(l x_c- \frac{x_c^2}{2}\right) - \cfrac{2\mu \sigma_y^0 x_c}{3\mu+H}\right]\,.
\end{eqnarray}
It is noted that $x_c$ is the transition point between the elastic and the elasto-plastic regions. There is no plastic deformation inside the bar when $x_c\leq 0$, implying
\begin{eqnarray}
	b_0 \leq \left(K+ \frac{4}{3}\mu\right)\frac{\sigma_y^0}{2\mu l}\,.
\end{eqnarray}

\bibliographystyle{elsarticle-num-names} 

\bibliography{reference}





\end{document}